\documentclass[11pt,a4paper]{article}

\PassOptionsToPackage{backref=page}{hyperref}
\usepackage{jheppub}
\usepackage[T1]{fontenc}
\usepackage{lmodern,microtype}
\usepackage{mathtools,bm}
\usepackage{xcolor}
\usepackage{xspace}

\newcommand{\bL}{\bar{L}}
\newcommand{\bh}{\bar{h}}
\newcommand{\bz}{\bar{z}}
\newcommand{\D}{\mathcal{D}}
\newcommand{\OO}{\mathcal{O}}
\newcommand{\MM}{\mathcal{M}}
\newcommand{\SSS}{\mathcal{S}}
\newcommand{\rad}{\operatorname{Rad}}
\newcommand{\dd}{\mathrm{d}}
\newcommand{\ii}{\mathrm{i}}
\newcommand{\Poincare}{Poincar\'{e}\xspace}

\numberwithin{equation}{section}

\definecolor{LinkColor}{rgb}{0.06,0.31,0.55}
\definecolor{CiteColor}{rgb}{0.0,0.55,0.0}
\definecolor{UrlColor}{rgb}{0.55,0.28,0.55}
\hypersetup{
    pdftitle={Massive Particles in Carrollian Holography},
    linktoc=all,
    colorlinks=true,
    linkcolor=LinkColor,
    citecolor=CiteColor,
    urlcolor=UrlColor
}

\title{Massive Particles in Carrollian Holography}
\author[a,b,c,d]{Bin Chen}
\author[c]{Zezhou Hu}
\author[e]{Reiko Liu}
\author[e,f]{Wen-Jie Ma}

\affiliation[a]{Institute of Fundamental Physics and Quantum Technology, \& School of Physical Science and Technology, Ningbo University, Ningbo, Zhejiang 315211, China}
\affiliation[b]{Zhejiang Key Laboratory of Extreme Universe, Ningbo University, Ningbo, Zhejiang 315211, China}
\affiliation[c]{School of Physics, Peking University, No.5 Yiheyuan Rd, Beijing 100871, China}
\affiliation[d]{Center for High Energy Physics, Peking University, No.5 Yiheyuan Rd, Beijing 100871, China}
\affiliation[e]{Shanghai Institute for Mathematics and Interdisciplinary Sciences (SIMIS), Shanghai 200433, China}
\affiliation[f]{Fudan Center for Mathematics and Interdisciplinary Study, Fudan University, Shanghai 200438, China}
\emailAdd{chenbin1@nbu.edu.cn}
\emailAdd{z.z.hu@pku.edu.cn}
\emailAdd{reiko.antoneva@foxmail.com}
\emailAdd{wenjie.ma@simis.cn}
\abstract{
    Understanding how massive particles are encoded in boundary operators is a central problem in Carrollian holography.
    We construct a three-dimensional Carrollian description of a massive scalar in four-dimensional flat spacetime, together with an explicit bulk-to-boundary integral transform.
    In this construction, nonzero mass requires infinitely many linearly independent primary states, while the absence of bulk spin imposes relations among their descendants.
    We show that these relations are equivalent to null-state decoupling and give a local constraint on the boundary fields.
    The integral transform preserves the full \Poincare symmetry and maps bulk scalar data to fields satisfying this constraint.
    We also obtain two- and three-point functions and the associated formal operator product expansion coefficients.
    The integral transform evaluates bulk data at complex momenta, making analytic continuation essential in Lorentzian signature.
This construction provides a concrete starting point for relating massive scattering amplitudes to Carrollian correlators and for reconstructing bulk fields from boundary data.}

\begin{document}
\maketitle

\section{Introduction}

One approach to flat-space holography is to express bulk scattering amplitudes as correlation functions of boundary operators.
This requires a dictionary relating bulk momentum-space data to fields with the appropriate boundary transformation laws.
In celestial holography~\cite{CheungDeLaFuenteSundrum2017,PasterskiShaoStrominger2017,PasterskiShao2017,Strominger2018,Raclariu2021,Pasterski2021,PasterskiPateRaclariu2021,McLoughlinPuhmRaclariu2022}, conformal primary wavefunctions provide such a dictionary, including for massive particles~\cite{PasterskiShaoStrominger2017,PasterskiShao2017,LawZlotnikov2020,IacobacciMuck2020,Narayanan2020,ChangMa2023,LiuMa2025}.
Carrollian holography retains a time coordinate on the null boundary and relates scattering amplitudes to Carrollian conformal correlators~\cite{Donnay2022,Bagchi2022,Nguyen2025}.
For massless particles, a modified Mellin transform provides an explicit relation between momentum-space amplitudes and boundary correlators~\cite{Banerjee2019}.
Highest-weight representations organize the boundary primary fields and their descendants, providing a framework for the corresponding local operator description~\cite{Donnay2022,Bagchi2022,NguyenWest2023,Nguyen2024}.

Extending this description to massive particles requires representations with a nonzero mass Casimir~\cite{ChenHu2024}, together with a map that preserves the bulk \Poincare transformations.\footnote{For earlier attempts and discussions on massive particles in Carrollian holography, see \emph{e.g.} \cite{ZhengDescendants2026,Zheng2026,Have2024,Melton2026}.}
Such a map is an intertwiner.
For scalar particles in a three-dimensional bulk, this approach has been used to construct a massive Carrollian basis with a two-dimensional boundary representation~\cite{Liu2026}.
Its integral kernel evaluates bulk data at complex momentum coordinates.

In this work, we construct the boundary representation from Carrollian primary states and their descendants and derive the corresponding intertwiner for a four-dimensional massive scalar.
With the primary-state conditions adopted here, it turns out that nonzero mass requires infinitely many linearly independent primary states.
The scalar transformation law requires the boundary fields produced by the map to have the same Casimir eigenvalues and to satisfy the Pauli--Lubanski condition.
In the family of representations considered here, the latter is stronger than Casimir matching alone.
We use it to determine the relations among descendants and show that imposing these relations is equivalent to removing null states.
This provides the local operator relations needed to construct correlation functions and to identify the boundary fields obtained from the integral transform.

The paper is organized as follows.
Section~\ref{sec:targetrequirements} defines the boundary representation as the image of a holographic map and derives the necessary Casimir and Pauli--Lubanski conditions from \Poincare covariance.
Section~\ref{sec:module} constructs the Carrollian module from its primary states and descendants, analyzes the Pauli--Lubanski condition, and identifies the null submodule whose removal gives the quotient representation.
Section~\ref{sec:local} develops the local operator realization of the quotient representation, derives its null-state equation, and constructs two- and three-point functions satisfying the Ward identities and null-state equations.
It also introduces the local realization of BPZ duality used in Appendix~\ref{app:ope} to extract OPE coefficients from the three-point function.
Section~\ref{sec:bulk} introduces the bulk massive scalar representation, formulates the \Poincare intertwining equations, and constructs and checks the integral kernel, specifying the analytic assumptions needed for the transform.
Section~\ref{sec:discussion} examines the kernel's support on the Lorentzian and Klein real slices and discusses how invertibility of the holographic map would lead to bulk reconstruction.
Appendix~\ref{app:formaldelta} explains the formal and holomorphic delta distributions used in the correlators and the bulk kernel, including the residue prescription for complex support.
Appendix~\ref{app:ope} derives the formal OPE coefficients from the three-point function using explicit Ward derivatives and BPZ pairings.

\section{Requirements on the boundary representation}
\label{sec:targetrequirements}

Our aim is to describe a four-dimensional massive scalar by fields on three-dimensional Carrollian spacetime.
Such a boundary description must preserve the \Poincare transformations of the bulk scalar.
The global Carrollian conformal algebra is isomorphic to the \Poincare algebra, but the transformation law of the boundary fields remains to be determined.
We must therefore find a boundary representation appropriate to a particle of mass $m>0$ and spin zero, together with a map from the bulk scalar to that representation.
In this section we derive necessary conditions on this boundary representation from the known bulk transformation law.

Let $\OO^{\rm B}(p)$ denote the bulk scalar operator in momentum space, with $p^\mu p_\mu=-m^2$ and $p^0>0$.
We write $\D_X^{\rm B}$ for the action of a \Poincare generator $X$ on $\OO^{\rm B}$.
Here $X$ is an element of the complexified \Poincare algebra $\mathfrak{iso}(1,3)$: a translation generator $P_\mu$, a Lorentz generator $M_{\mu\nu}$, or a linear combination of them.
The superscript $\mathrm{B}$ denotes the bulk.
For a scalar, translations act by multiplication by momentum and Lorentz transformations by derivatives with respect to momentum, as given explicitly in \eqref{eq:targetbulkorbital} below.

We seek a nonzero linear holographic map $K$ from the bulk operator to a boundary Carrollian field,
\begin{equation*}
    \OO^{\rm b}=K\OO^{\rm B}
    \, ,
\end{equation*}
where the superscript $\mathrm{b}$ denotes the boundary.
In the following, $\D_X^{\rm b}$ denotes the action of $X$ on $\OO^{\rm b}$.
Though the action $\D_X^{\rm B}$ is known, $\D_X^{\rm b}$ and $K$ are to be determined.

The bulk and boundary descriptions are dual to each other.
A translation or Lorentz transformation of the bulk scalar must therefore have a corresponding action on its boundary description.
Here the bulk \Poincare generators are identified with the generators of the global Carrollian conformal algebra.
Although their actions on bulk and boundary fields take different forms, they represent the same physical symmetry.
We accordingly require the holographic map $K$ to preserve this correspondence between symmetry transformations.
Transforming a bulk field and then applying $K$ must give the same result as applying $K$ first and transforming the resulting boundary field.
We call this requirement \emph{\Poincare covariance} of the holographic map.
Infinitesimally, it reads
\begin{equation}
    \D_X^{\rm b}\OO^{\rm b}
    =\D_X^{\rm b}K\OO^{\rm B}
    =K\D_X^{\rm B}\OO^{\rm B}
    \, ,
    \qquad X\in\mathfrak{iso}(1,3)\, .
    \label{eq:targetintertwining}
\end{equation}
Equation~\eqref{eq:targetintertwining} is referred to as the \emph{intertwining equation}.
We impose it as a physical consistency condition on the proposed holographic map; it constrains both $K$ and the boundary transformation law.
A linear map satisfying this equation is called an \emph{intertwiner}.
We take the domain of $K$ to be closed under the bulk generator actions and require \eqref{eq:targetintertwining} to hold throughout this domain.
Its image $\operatorname{Im}K$ is then closed under the boundary actions and carries the boundary representation realized by $K$.

The symmetry action on a physical system also determines successive transformations generated by ordered products of the generators.
It therefore extends to the universal enveloping algebra, which contains the Casimir operators and the Pauli--Lubanski vector.
We now specify how products of symmetry generators act.
The universal enveloping algebra $U(\mathfrak{g})$ of the complexified \Poincare algebra $\mathfrak{g}=\mathfrak{iso}(1,3)$ consists of finite linear combinations of ordered products of generators and the identity, subject to $XY-YX=[X,Y]$.
In our field convention $[X,\OO^a]=\D_X^a\OO^a$, where $a=\mathrm{b},\mathrm{B}$, the differential operators act on the field's coordinate dependence and satisfy $[\D_X^a,\D_Y^a]=-\D_{[X,Y]}^a$.
Their extension to $U(\mathfrak{g})$ therefore reverses the order of each product:
\begin{equation*}
    \D_{X_1\cdots X_n}^a
    =\D_{X_n}^a\cdots\D_{X_1}^a
    \, .
\end{equation*}
We extend this rule by linearity, with the identity element acting as the identity operator.
It describes successive symmetry transformations: for example, $\D_{XY}^a\OO^a=[X,[Y,\OO^a]]$, rather than $[XY,\OO^a]$.

The intertwining equation~\eqref{eq:targetintertwining} then extends to $U(\mathfrak{g})$.
We require this equation to hold on the bulk fields and all their successive transforms specified above.
For $X,Y\in\mathfrak{g}$, applying it twice gives
\begin{equation}
    \D_Y^{\rm b}\D_X^{\rm b}\OO^{\rm b}
    =\D_Y^{\rm b}K\D_X^{\rm B}\OO^{\rm B}
    =K\D_Y^{\rm B}\D_X^{\rm B}\OO^{\rm B}
    \, .
    \label{eq:targetproducttransfer}
\end{equation}
The second equality applies \eqref{eq:targetintertwining} to $\D_X^{\rm B}\OO^{\rm B}$, which lies in the chosen domain of $K$.
Repeating this argument for longer products and using linearity yields
\begin{equation}
    \D_A^{\rm b}\OO^{\rm b}
    =\D_A^{\rm b}K\OO^{\rm B}
    =K\D_A^{\rm B}\OO^{\rm B}
    \, ,
    \qquad A\in U(\mathfrak{g})\, .
    \label{eq:targetenveloping}
\end{equation}
This relation holds for every $A\in U(\mathfrak{g})$, including noncentral elements such as the components of the Pauli--Lubanski vector.

The \Poincare Casimirs are elements of $U(\mathfrak{g})$ that commute with every generator.
For a massive particle, the two independent Casimirs encode its mass and spin.
With $\eta_{\mu\nu}=\operatorname{diag}(-1,1,1,1)$ and $\epsilon^{0123}=1$, our conventions are
\begin{equation}
    C_{\rm mass}=P^\mu P_\mu
    \, ,\qquad
    C_{\rm spin}=W^\mu W_\mu
    \, ,
    \label{eq:targetcasimirs}
\end{equation}
where $W^\mu$ is the Pauli--Lubanski vector, constructed from the translation and Lorentz generators
\begin{equation}
    W^\mu=\frac{1}{2}\epsilon^{\mu\nu\rho\sigma}M_{\nu\rho}P_\sigma
    \, ,
\end{equation}
with repeated spacetime indices summed and indices raised and lowered with $\eta_{\mu\nu}$.
For a scalar, the Lorentz generators act purely orbitally.
In momentum coordinates the bulk generators act as
\begin{equation}
    \begin{aligned}
    \D_{P_\sigma}^{\rm B}\OO^{\rm B}
    &=\ii s p_\sigma\OO^{\rm B}
    \, ,\qquad s=\pm1\, ,\\
    \D_{M_{\nu\rho}}^{\rm B}\OO^{\rm B}
    &=\left(p_\rho\frac{\partial}{\partial p^\nu}
    -p_\nu\frac{\partial}{\partial p^\rho}\right)\OO^{\rm B}
    \, .
    \end{aligned}
    \label{eq:targetbulkorbital}
\end{equation}
In Minkowski space, we take $p^\mu$ on the future mass shell, $p^0>0$, and use $s=+1$ ($s=-1$) for outgoing (incoming) operators.\footnote{
    Alternatively, in Minkowski space one may absorb $s$ into $p^\mu$ by allowing $p^\mu$ to range over both disconnected components of the real mass shell, namely the positive-energy branch $p^0>0$ and the negative-energy branch $p^0<0$.
In Klein signature $(2,2)$, the real massive shell is connected, so a Lorentz transformation can reverse the sign of the energy, and the sign $s$ is normally absorbed into $p^\mu$.}
Although written in terms of four-momentum derivatives, the Lorentz vector fields preserve $p^2$ and therefore act within the mass shell.
The translation action gives $\D_{C_{\rm mass}}^{\rm B}\OO^{\rm B}=-p^2\OO^{\rm B}=m^2\OO^{\rm B}$.
The positive Casimir eigenvalue comes from the factor $\ii$ in \eqref{eq:targetbulkorbital}.

With the above actions, the definition of the Pauli--Lubanski vector gives
\begin{equation}
    \D_{W^\mu}^{\rm B}\OO^{\rm B}
    =\frac{\ii s}{2}\epsilon^{\mu\nu\rho\sigma}p_\sigma
    \left(p_\rho\frac{\partial}{\partial p^\nu}
    -p_\nu\frac{\partial}{\partial p^\rho}\right)\OO^{\rm B}=0
    \, .
    \label{eq:targetbulkPL}
\end{equation}
Here we have used the fact that antisymmetry of the Levi--Civita tensor gives $\epsilon^{\mu\nu\rho\sigma}p_\sigma p_\rho=0$ and $\epsilon^{\mu\nu\rho\sigma}p_\sigma p_\nu=0$.
Thus every component of $W^\mu$ annihilates the bulk scalar, and its spin Casimir vanishes as well.
This is a property of the transformation law of the scalar operator $\OO^{\rm B}$.
The boundary fields produced by $K$ inherit this property through \eqref{eq:targetenveloping}.
Taking $A=W^\mu$ gives
\begin{equation}
    \D_{W^\mu}^{\rm b}\OO^{\rm b}
    =K\D_{W^\mu}^{\rm B}\OO^{\rm B}=0
    \, .
    \label{eq:targetPLtransfer}
\end{equation}
The last equality follows from \eqref{eq:targetbulkPL}.

Equation~\eqref{eq:targetenveloping} also transfers the actions of the two Casimir operators.
Hence every boundary field $\OO^{\rm b}$ in the image of $K$ must satisfy
\begin{equation}
    \D_{C_{\rm mass}}^{\rm b}\OO^{\rm b}=m^2\OO^{\rm b}
    \, ,\qquad
    \D_{C_{\rm spin}}^{\rm b}\OO^{\rm b}=0
    \, ,\qquad
    \D_{W^\mu}^{\rm b}\OO^{\rm b}=0\quad(\mu=0,1,2,3)
    \, .
    \label{eq:targetrequirements}
\end{equation}
We call the first two equations the \emph{mass Casimir condition} and the \emph{spin Casimir condition}, respectively.
Together they express \emph{Casimir matching}: the boundary representation has the same mass and spin Casimir eigenvalues as the bulk scalar.
The \emph{Pauli--Lubanski condition} is the last equation, requiring all four components $W^\mu$ to annihilate every field in the boundary representation.
It implies the spin Casimir condition, but is stronger in the class of representations considered here.
We retain both Casimir conditions because they first constrain the Carrollian construction in Section~\ref{sec:module}.
There we suppress the differential-action notation and write the corresponding representation conditions as $C_{\rm mass}=m^2\mathbf{1}$, $C_{\rm spin}=0$ and $W^\mu=0$.

The requirements in \eqref{eq:targetrequirements} specify the properties that a boundary representation must have to describe the bulk massive scalar.
In Section~\ref{sec:module}, we construct such a representation by first matching the two Casimirs and then imposing the stronger Pauli--Lubanski condition.
We show that, at the selected scaling dimension $\Delta=1$, the latter is equivalent to quotienting by the null states.

\section{Constructing the Carrollian representation}
\label{sec:module}

\subsection{Primary states and descendants}

The global Carrollian conformal algebra is the complexified \Poincare algebra $\mathfrak{g}=\mathfrak{iso}(1,3)$.
In a boundary-adapted basis its nonzero brackets are
\begin{align}
    [L_n,L_m]&=(n-m)L_{n+m},&
    [\bL_n,\bL_m]&=(n-m)\bL_{n+m},\notag\\
    [L_n,P_{k,l}]&=\left(\frac{n}{2}-k\right)P_{n+k,l},&
    [\bL_n,P_{k,l}]&=\left(\frac{n}{2}-l\right)P_{k,n+l}
    \, ,
    \label{eq:algebra}
\end{align}
where $n,m=-1,0,1$ and $k,l=\pm\tfrac{1}{2}$.
The two $\mathfrak{sl}(2)$ subalgebras commute with each other, and all translation generators commute.
We write $L_\pm=L_{\pm1}$, which are related to the generators of $\mathfrak{g}=\mathfrak{iso}(1,3)$ by
\begin{align}
    L_-&=\tfrac{1}{2}(M_{01}+M_{13}-\ii M_{02}-\ii M_{23}),&
    \bL_-&=\tfrac{1}{2}(M_{01}+M_{13}+\ii M_{02}+\ii M_{23}),\notag\\
    L_0&=\tfrac{1}{2}(-M_{03}+\ii M_{12}),&
    \bL_0&=\tfrac{1}{2}(-M_{03}-\ii M_{12}),\notag\\
    L_+&=\tfrac{1}{2}(-M_{01}+M_{13}-\ii M_{02}+\ii M_{23}),&
    \bL_+&=\tfrac{1}{2}(-M_{01}+M_{13}+\ii M_{02}-\ii M_{23})
    \, ,
    \label{eq:dictionaryLorentz}
\end{align}
where the $M_{\mu\nu}$ are generators of the Lorentz group.
We reorganize the translations by defining
\begin{equation}
    P_{--}=P_0-P_3
    \, ,\quad P_{+-}=-(P_1+\ii P_2)\, ,\quad
    P_{-+}=-(P_1-\ii P_2)
    \, ,\quad P_{++}=P_0+P_3\, ,
    \label{eq:dictionaryP}
\end{equation}
where the $P_\mu$ are generators of translations.

The construction of the module is guided by a nonzero eigenvalue of the mass Casimir.
We seek a module on which
\begin{equation}
    C_{\rm mass}=-P_{--}P_{++}+P_{+-}P_{-+}=\xi^2\mathbf{1}
    \, ,
    \qquad \xi\ne0\, .
    \label{eq:masscasimir}
\end{equation}
We organize states in a module by the eigenvalue of $L_0+\bL_0$: a state with weights $(h,\bh)$ under $(L_0,\bL_0)$ has eigenvalue $h+\bh$.
The \emph{raising generators} $L_-,\bL_-,P_{--}$ raise it by one, while the \emph{lowering generators} $L_+,\bL_+,P_{++}$ lower it by one.
The primary space is defined to be annihilated by the lowering generators and is preserved by the operators $L_0,\bL_0,P_{+-},P_{-+}$.
On this space, the annihilation of $P_{++}$ reduces the action of \eqref{eq:masscasimir} to $P_{+-}P_{-+}=\xi^2\mathbf{1}$, so the two commuting translations are invertible, each with inverse $\xi^{-2}$ times the other.

The Cartan commutators in \eqref{eq:algebra} show that $P_{+-}$ and $P_{-+}$ shift the joint weights by $(-\tfrac{1}{2},\tfrac{1}{2})$ and $(\tfrac{1}{2},-\tfrac{1}{2})$, respectively.
Thus a one-dimensional primary space would force both translations to vanish and could only have zero mass Casimir.
For $\xi\ne0$, their invertibility instead requires a weight orbit extending in both directions.
We choose the orbit of one joint eigenvector $e_0$ of weights $(h,\bar{h})$, with one-dimensional weight spaces, and normalize successive states so that both translation coefficients equal $\xi$.
This gives the algebraic direct sum $E_{h,\bar{h},\xi}=\bigoplus_{n\in\mathbb{Z}}\mathbb{C} e_n$ with action
\begin{align}
    L_0e_n&=\left(h+\frac{n}{2}\right)e_n,&
    \bL_0e_n&=\left(\bar{h}-\frac{n}{2}\right)e_n,\notag\\
    P_{+-}e_n&=\xi e_{n-1},&P_{-+}e_n&=\xi e_{n+1},\notag\\
    L_+e_n&=\bL_+e_n=P_{++}e_n=0, &n\in \mathbb{Z}
    \, .
    \label{eq:seed}
\end{align}
This primary module is infinite-dimensional.
It is irreducible as well: polynomial interpolation in $L_0$ isolates a nonzero component of any vector with finite support, and repeated translations then generate every $e_n$.
For later convenience, we set $\Delta=h+\bar{h}$ and $J=h-\bar{h}$.

We now generate all descendants by acting with $L_-,\bL_-,P_{--}$, without imposing further relations.
Let $\mathfrak{p}$ be the subalgebra generated by $L_0,\bL_0,P_{+-},P_{-+},L_+,\bL_+,P_{++}$.
The resulting Carrollian module is
\begin{equation}
    \MM_{h,\bar{h},\xi}
    =U(\mathfrak{g})\otimes_{U(\mathfrak{p})}E_{h,\bar{h},\xi}
    =\operatorname{span}\{L_-^a\bL_-^bP_{--}^ce_n\}
    \, ,
    \quad a,b,c\in \mathbb{N}\, ,\quad n\in\mathbb{Z}\, .
    \label{eq:induction}
\end{equation}
The three raising generators commute, and the level of a \Poincare--Birkhoff--Witt (PBW) basis monomial is $a+b+c$.
Throughout, vectors in $\MM_{h,\bar{h},\xi}$ are finite linear combinations of these descendants.
Equation~\eqref{eq:seed} gives $C_{\rm mass}e_n=\xi^2e_n$; centrality then extends this eigenvalue to all of $\MM_{h,\bar{h},\xi}$.
The mass condition leaves $h$ and $\bar{h}$ unrestricted.

The second Casimir acts on $\MM_{h,\bar{h},\xi}$ as
\begin{equation}
    C_{\rm spin}=\xi^2(\Delta-1)(\Delta-2)\mathbf{1}
    \, .
    \label{eq:inducedspincasimir}
\end{equation}
This follows by applying the Cartesian expression for $W^\mu W_\mu$ to \eqref{eq:seed}; centrality extends the result to all descendants.
Matching both Casimirs to the bulk scalar values in \eqref{eq:targetrequirements} therefore requires
\begin{equation}
    \xi^2=m^2
    \, ,\qquad \Delta=1\ \hbox{or}\ 2\, .
    \label{eq:casimircandidates}
\end{equation}
The Pauli--Lubanski condition $W^\mu=0$ is stronger than $C_{\rm spin}=0$.
We will discuss it in the next subsection.

\subsection{The Pauli--Lubanski condition}\label{subsec:scalarquotient}

To impose the Pauli--Lubanski condition derived in Section~\ref{sec:targetrequirements}, introduce the level-one vectors
\begin{equation}
    R_n^{(J)}=\xi\bL_-e_{n+1}-\xi L_-e_{n-1}
    +(n+J)P_{--}e_n
    \, .
    \label{eq:null}
\end{equation}
The Cartesian expression for $W^\mu$ and \eqref{eq:seed} give
\begin{align}
    W^0e_n&=W^3e_n=\frac{\ii}{2}R_n^{(J)},\notag\\
    W^1e_n&=\frac{\ii\xi}{2}(1-\Delta)(e_{n-1}-e_{n+1}),\notag\\
    W^2e_n&=\frac{\xi}{2}(1-\Delta)(e_{n-1}+e_{n+1})
    \, .
    \label{eq:plseed}
\end{align}
Equation~\eqref{eq:plseed} holds in $\MM_{h,\bar{h},\xi}$, where the vectors on its right-hand side need not vanish.
Consider a quotient representation satisfying \eqref{eq:targetrequirements}, and let $\pi$ denote the projection from $\MM_{h,\bar{h},\xi}$ onto it.
Since $\pi$ preserves the generator actions and hence their products, the Pauli--Lubanski condition requires
\begin{equation}
    0=W^\mu\pi(e_n)=\pi(W^\mu e_n)
    \, .
    \label{eq:PLquotientconstraint}
\end{equation}
For $\Delta\ne1$, the combinations $W^1\pm\ii W^2$ in \eqref{eq:plseed} give nonzero multiples of $e_{n\mp1}$, so \eqref{eq:PLquotientconstraint} sets every $\pi(e_n)$ to zero.
As the projected primary states generate the quotient, it is zero.
Thus the $\Delta=2$ branch allowed by Casimir matching is excluded.

At $\Delta=1$, $W^1e_n=W^2e_n=0$, while the remaining components are proportional to $R_n^{(J)}$.
Equation~\eqref{eq:PLquotientconstraint} therefore requires $\pi(R_n^{(J)})=0$.
Since $\ker\pi$ is a submodule, it contains $\mathcal{N}=U(\mathfrak{g})\{R_n^{(J)}\}$, including all symmetry descendants of these vectors.
Every such quotient thus factors through
\begin{equation}
    \SSS_{J,\xi}=\MM_{(1+J)/2,(1-J)/2,\xi}/\mathcal{N}
    \, .
    \label{eq:quotient}
\end{equation}

It remains to extend $W^\mu=0$ from the primary states to all of $\SSS_{J,\xi}$.
The Pauli--Lubanski vector commutes with the translations and transforms as a vector under Lorentz transformations:
\begin{equation}
    [P_\alpha,W^\mu]=0
    \, ,\qquad
    [M_{\alpha\beta},W^\mu]
    =\delta_\beta^{\mu}W_\alpha-\delta_\alpha^{\mu}W_\beta
    \, .
    \label{eq:PLcovariance}
\end{equation}
Here $W_\alpha=\eta_{\alpha\nu}W^\nu$.
Thus $[W^\mu,X]$ is a linear combination of the $W^\nu$ for every $X\in\mathfrak{g}$.
If $W^\nu v\in\mathcal{N}$ for all $\nu$, then
\begin{equation}
    W^\mu(Xv)=X(W^\mu v)+[W^\mu,X]v\in\mathcal{N}
    \, ,
    \label{eq:PLdescendant}
\end{equation}
because $\mathcal{N}$ is a submodule.
At $\Delta=1$, \eqref{eq:plseed} supplies the initial step $W^\mu e_n\in\mathcal{N}$.
Induction on the number of raising generators therefore gives $W^\mu\MM_{(1+J)/2,(1-J)/2,\xi}\subseteq\mathcal{N}$, proving $W^\mu=0$ on $\SSS_{J,\xi}$.
Hence quotienting by $\mathcal{N}$ imposes precisely the Pauli--Lubanski condition.
We next show that $\mathcal{N}$ is the full BPZ radical and that this quotient is nonzero and irreducible.

\subsection{Null states and the quotient representation}
\label{subsec:radical}

The complex-linear BPZ anti-involution is $L_n^\dagger=L_{-n}$, $\bL_n^\dagger=\bL_{-n}$ and $P_{k,l}^\dagger=P_{-k,-l}$.
It defines a symmetric contravariant form by
\begin{equation}
    \hat{B}(e_n,e_m)=B_0\delta_{nm}
    \, ,\qquad
    \hat{B}(Xv,w)=\hat{B}(v,X^\dagger w)
    \, ,\qquad B_0\ne0\, .
    \label{eq:bpz}
\end{equation}
We define null states to be elements of the radical $\rad \hat{B}$, namely vectors $v$ satisfying $\hat{B}(v,w)=0$ for every $w$ in the module.
Imposing the \emph{null-state condition} means passing to the quotient by $\rad\hat{B}$, so that all null states are identified with zero.

At level one, consider the basis $(L_-e_{r-1},\bL_-e_{r+1},P_{--}e_r)$ of the subspace with eigenvalue $J+r$ under $L_0-\bL_0$.
Contravariance and the primary-state conditions give the following Gram matrix on $\MM_{h,\bar{h},\xi}$, before taking the quotient:
\begin{equation}
    \mathsf{B}^{\mathcal{M}}_{1,r}=B_0
    \begin{pmatrix}
        2h+r-1&0&\xi\\
        0&2\bar{h}-r-1&\xi\\
        \xi&\xi&0
    \end{pmatrix}
    \, ,\qquad
    \det \mathsf{B}^{\mathcal{M}}_{1,r}=-2B_0^3\xi^2(\Delta-1)
    \, .
    \label{eq:gram1}
\end{equation}
At $\Delta=1$, the coefficient vector of $R_n^{(J)}$ in \eqref{eq:null} lies in the kernel of $\mathsf{B}^{\mathcal{M}}_{1,n}$.
Each $R_n^{(J)}$ is annihilated by all three lowering generators and is therefore a singular vector.
Contravariance then implies that their generated submodule $\mathcal{N}$ lies in $\rad\hat{B}$.

We write $[v]=v+\mathcal{N}$ for the equivalence class of $v$ in $\SSS_{J,\xi}$.
Since $\mathcal{N}\subseteq\rad\hat{B}$, the bilinear form descends to the quotient, with $\hat{B}([v],[w])=\hat{B}(v,w)$; we use the same notation for the induced form.

To establish that $\mathcal{N}$ is the full radical, use the quotient relation
\begin{equation}
    [\bL_-e_n]=[L_-e_{n-2}]
    -\frac{n+J-1}{\xi}[P_{--}e_{n-1}]
    \label{eq:reduction}
\end{equation}
to eliminate powers of $\bL_-$.
At level $\ell$, the subspace with eigenvalue $J+r$ under $L_0-\bL_0$ is therefore spanned by
\begin{equation}
    v_{\ell r,a}\coloneqq[L_-^aP_{--}^{\ell-a}e_{r-a}]
    \, ,\qquad 0\le a\le\ell\, .
    \label{eq:quotientbasis}
\end{equation}
We define the Gram matrix $\mathsf{B}_{\ell,r}$ of this spanning family in the quotient $\SSS_{J,\xi}$ by its entries,
\begin{equation*}
    (\mathsf{B}_{\ell,r})_{ab}
    \coloneqq\hat{B}(v_{\ell r,a},v_{\ell r,b})
    \, ,
    \qquad 0\le a,b\le\ell\, .
\end{equation*}
Its entries obey
\begin{equation}
    (\mathsf{B}_{\ell,r})_{ab}=0\quad(a+b<\ell)
    \, ,\qquad
    (\mathsf{B}_{\ell,r})_{a,\ell-a}=B_0\,a!(\ell-a)!\,\xi^\ell
    \, .
    \label{eq:antidiagonal}
\end{equation}
To evaluate the pairing, commute $P_{++}^{\ell-a}$ through the $b$ factors of $L_-$ on the right.
If $b<\ell-a$, at least one factor of $P_{++}$ annihilates the primary state.
For $b=\ell-a$, repeated use of $[P_{++},L_-]=P_{-+}$ and $[L_+,P_{--}]=P_{+-}$ gives the factorials in \eqref{eq:antidiagonal}, while the translations contribute $\xi^\ell$.
Consequently
\begin{equation}
    \det \mathsf{B}_{\ell,r}
    =(-1)^{\ell(\ell+1)/2}B_0^{\ell+1}
    \xi^{\ell(\ell+1)}\left(\prod_{a=0}^{\ell}a!\right)^2\ne0
    \, .
    \label{eq:quotientdet}
\end{equation}
The nonzero determinant establishes that the vectors in \eqref{eq:quotientbasis} are linearly independent and that the quotient form is nondegenerate on every joint weight space.
Thus $\mathcal{N}=\rad \hat{B}$.
The quotient is also algebraically irreducible.
Any nonzero submodule contains a nonzero joint weight vector.
Nondegeneracy implies that a polynomial in the lowering generators maps this vector to a nonzero primary state.
Irreducibility of the primary module and the action of the raising generators then generate the entire quotient.

At $\Delta=1$, the Pauli--Lubanski condition is therefore equivalent to the null-state condition: the relations derived in Section~\ref{subsec:scalarquotient} remove exactly the full radical.
The resulting $\SSS_{J,\xi}$ is a nonzero irreducible quotient satisfying the conditions in \eqref{eq:targetrequirements} once $\xi^2=m^2$ is imposed.

The condition $W^\mu=0$ leaves $J$ unrestricted.
Integer shifts of $J$ can be absorbed by relabeling the primary states.
In the following discussion, we focus on the symmetric choice
\begin{equation}
    J=0
    \, ,\qquad h=\bar{h}=\frac{1}{2}\, ,\qquad \SSS_{0,\xi}\, .
    \label{eq:scalarweights}
\end{equation}
We next describe the local operator realization of this quotient, and then construct a holographic map obeying its local relations.

\section{Local operators and correlation functions}
\label{sec:local}

\subsection{Local operator realization and the null-state equation}
\label{subsec:localrealization}

We now construct the local operator realization of the quotient $\SSS_{0,\xi}=\MM_{1/2,1/2,\xi}/\mathcal{N}$ obtained in Section~\ref{sec:module}.
Let $(u,z,\bz)$ be local coordinates on the three-dimensional Carrollian spacetime, where $u$ is Carrollian time and $z,\bz$ are algebraically independent coordinates on the complexified spatial slice.
The boundary primary states are the quotient classes
\begin{equation*}
    \lvert\OO_n\rangle\coloneqq[e_n]\in\SSS_{0,\xi}
    \, .
\end{equation*}
Their formal generating family is
\begin{equation}
    \lvert\OO(t)\rangle
    \coloneqq\sum_{n\in\mathbb{Z}}t^n\lvert\OO_n\rangle
    =\sum_{n\in\mathbb{Z}}t^n[e_n]
    \, .
    \label{eq:generatingstate}
\end{equation}
Here $t$ is an auxiliary variable whose powers mark the components $\lvert\OO_n\rangle$, and the sum is understood coefficientwise.

The corresponding local operators form the generating field
\begin{equation*}
    \OO^{\rm b}(u,z,\bz,t)
    =\sum_{n\in\mathbb{Z}}t^n\OO_n^{\rm b}(u,z,\bz)
    \, .
\end{equation*}
At the spacetime origin, the state--operator correspondence reads
\begin{equation*}
    \OO_n^{\rm b}(0,0,0)\lvert0\rangle
    =\lvert\OO_n\rangle
    \, ,\qquad
    \OO^{\rm b}(0,0,0,t)\lvert0\rangle
    =\lvert\OO(t)\rangle
    \, .
\end{equation*}
Here $\lvert0\rangle$ is the symmetry-invariant vacuum, and the second equality is also understood coefficientwise in $t$.
Hereafter, $0$ denotes the spacetime origin $u=z=\bz=0$.
The fields away from the origin are obtained by translation with $U=uP_{--}+zL_-+\bz\bL_-$:
\begin{equation}
    \OO_n^{\rm b}(u,z,\bz)=e^U\OO_n^{\rm b}(0)e^{-U}
    \, ,\qquad
    \OO^{\rm b}(u,z,\bz,t)=e^U\OO^{\rm b}(0,t)e^{-U}
    \, .
    \label{eq:local}
\end{equation}

The generator action is inherited from $\MM_{1/2,1/2,\xi}$ through $X\lvert\OO_n\rangle=[Xe_n]$.
Descendants generated by $L_-,\bL_-,P_{--}$ therefore correspond to coordinate derivatives of these operators at the origin.
The differential transformation laws below can be computed using the representatives $e_n$ in $\MM_{1/2,1/2,\xi}$.
To realize $\SSS_{0,\xi}$, the relations $[R_n^{(0)}]=0$ and their symmetry descendants must also be imposed as operator identities.
We derive the resulting local null-state equation below.

Define the differential action by $[X,\OO^{\rm b}]=\D_X^{\rm b}\OO^{\rm b}$.
Then $[\D_X^{\rm b},\D_Y^{\rm b}]=-\D_{[X,Y]}^{\rm b}$, since these operators act on the coordinate dependence of a translated operator.
The Baker--Campbell--Hausdorff conjugation using \eqref{eq:seed} gives
\begin{align}
    \D_{L_-}^{\rm b}&=\partial_z,&\D_{\bL_-}^{\rm b}&=\partial_{\bz},&
    \D_{P_{--}}^{\rm b}&=\partial_u,\notag\\
    \D_{L_0}^{\rm b}&=z\partial_z+\frac{u}{2}\partial_u+\frac{1}{2}+\frac{t}{2}\partial_t,&
    \D_{\bL_0}^{\rm b}&=\bz\partial_{\bz}+\frac{u}{2}\partial_u+\frac{1}{2}-\frac{t}{2}\partial_t
    \, ,
    \label{eq:localcartan}
\end{align}
and
\begin{align}
    \D_{P_{+-}}^{\rm b}&=z\partial_u+\xi t,&
    \D_{P_{-+}}^{\rm b}&=\bz\partial_u+\xi t^{-1},\notag\\
    \D_{L_+}^{\rm b}&=z^2\partial_z+zu\partial_u+z+zt\partial_t+\xi ut,&
    \D_{\bL_+}^{\rm b}&=\bz^2\partial_{\bz}+\bz u\partial_u+\bz-\bz t\partial_t+\xi ut^{-1},
    \notag\\
    \D_{P_{++}}^{\rm b}&=z\bz\partial_u+\xi\bz t+\xi zt^{-1}
    \, .
    \label{eq:locallowering}
\end{align}
For example, $\sum_n t^n\OO_{n-1}^{\rm b}=t\OO^{\rm b}$ expresses the shift of the component index as multiplication by $t$.
In terms of the generating function, the null-state relations become
\begin{equation}
    Q_\xi\OO^{\rm b}=0
    \, ,\qquad
    Q_\xi=\xi\partial_{\bz}-\xi t^2\partial_z
    +t^2\partial_u\partial_t
    \, .
    \label{eq:quotientPDE}
\end{equation}
We stress that the differential action above is that of $\MM_{1/2,1/2,\xi}$; imposing the null-state equation \eqref{eq:quotientPDE} in addition implements the quotient constructed in Section~\ref{sec:module}.

The relation to the Pauli--Lubanski condition is explicit.
Using the reversed product order $\D_{XY}^{\rm b}=\D_Y^{\rm b}\D_X^{\rm b}$, the Cartesian dictionary and \eqref{eq:localcartan}--\eqref{eq:locallowering} give
\begin{align}
    W^0+W^3
    &=\ii\bigl[(L_0-\bL_0)P_{--}-L_-P_{+-}+\bL_-P_{-+}\bigr],
    \notag\\
    \D_{W^0+W^3}^{\rm b}
    &=\ii\bigl(t\partial_t\partial_u-\xi t\partial_z
    +\xi t^{-1}\partial_{\bz}\bigr)
    =\frac{\ii}{t}Q_\xi
    \, .
    \label{eq:PLnullbridge}
\end{align}
This identity holds for the differential action of $\MM_{1/2,1/2,\xi}$, before imposing any null-state relation.
Thus the Pauli--Lubanski condition implies $Q_\xi\OO^{\rm b}=0$.
Conversely, comparing the coefficients of $t^{n+1}$ in this equation gives
\begin{equation}
    \xi\partial_{\bz}\OO_{n+1}^{\rm b}
    -\xi\partial_z\OO_{n-1}^{\rm b}+n\partial_u\OO_n^{\rm b}=0
    \, .
    \label{eq:quotientcomponents}
\end{equation}
At the origin these set the images of $R_n^{(0)}$ to zero under $e_n\mapsto\OO_n^{\rm b}(0)$.
An equivariant state-to-operator map then kills their symmetry descendants and factors through $\SSS_{0,\xi}$.
By Section~\ref{subsec:scalarquotient}, every $W^\mu$ vanishes on this quotient.
Together with Section~\ref{subsec:radical}, this establishes the equivalence of the Pauli--Lubanski and null-state conditions in this covariant local realization at $\Delta=1$ and $J=0$.

\subsection{BPZ duality}
\label{subsec:bpz}

The state--operator correspondence also relates the BPZ duals of states to outgoing operator insertions.
For $v,w\in\SSS_{0,\xi}$, define the linear functional ${}_{\rm BPZ}\langle w\rvert$ by
\begin{equation}
    {}_{\rm BPZ}\langle w\vert v\rangle=\hat{B}(v,w)
    \, ,\qquad
    {}_{\rm BPZ}\langle yw\rvert
    ={}_{\rm BPZ}\langle w\rvert y^\dagger
    \, ,
    \quad y\in U(\mathfrak{g})\, .
    \label{eq:opeBPZdual}
\end{equation}
The generator action is inherited from $\MM_{1/2,1/2,\xi}$, and $\dagger$ is the BPZ anti-involution defined in \eqref{eq:bpz}.

To realize BPZ duality on local fields, we require an inversion $\mathcal{I}_\xi$ to obey
\begin{equation}
    \mathcal{I}_\xi\D_X^{\rm b}
    =-\D_{X^\dagger}^{\rm b}\mathcal{I}_\xi
    \, ,
    \qquad X\in\mathfrak{g}\, .
    \label{eq:bpzrequirement}
\end{equation}
A local realization of this requirement on the punctured boundary chart is
\begin{equation}
    (\mathcal{I}_\xi\OO^{\rm b})(u,z,\bz,t)
    =\frac{e^{-\xi u(t/z+1/(\bz t))}}{z\bz}\,
    \OO^{\rm b}\!\left(-\frac{u}{z\bz},\frac{1}{z},
    \frac{1}{\bz},\frac{t\bz}{z}\right)
    \, .
    \label{eq:opeBPZinversion}
\end{equation}
The inversion \eqref{eq:opeBPZinversion} satisfies the BPZ relation \eqref{eq:bpzrequirement} for all the \Poincare generators, with their differential actions given in \eqref{eq:localcartan}--\eqref{eq:locallowering}.
This realization also preserves the null-state equation.
Whereas $\OO_n^{\rm b}(0)\lvert0\rangle=\lvert\OO_n\rangle$ creates a primary state, its outgoing dual ${}_{\rm BPZ}\langle\OO_n\rvert$ is obtained from $\langle0\rvert(\mathcal{I}_\xi\OO^{\rm b})(x,t)$ by selecting the $n$-th auxiliary-variable coefficient and the primary coefficient at the origin of the inverted chart.
This operation is performed on the transformed correlator with its coefficient prescription; it is not a substitution of zero into the individual inverse coordinates in \eqref{eq:opeBPZinversion}.
Descendant duals follow from $y^\dagger$ in \eqref{eq:opeBPZdual}.
The two-point calculation below fixes the normalization, and Appendix~\ref{app:ope} applies this construction to three-point data and the OPE.

\subsection{Two-point functions}
\label{subsec:twopoint}

For two operators in the same representation, let $u=u_1-u_2$, $z=z_1-z_2$ and $\bz=\bz_1-\bz_2$, and restrict to $z\bz\ne0$.
The two-point function determined by the Ward identities takes the form
\begin{equation}
    G(1,2)=\frac{C}{z\bz}
    \exp\!\left[-\frac{\xi u(t_1+t_2)}{z}\right]
    \delta^\times\!\left(\frac{\bz t_1t_2}{z}\right)
    \, ,\qquad
    \delta^\times(x)=\sum_{r\in\mathbb{Z}}x^r
    \, .
    \label{eq:twopoint}
\end{equation}
Here $\delta^\times$ is the formal delta distribution, satisfying $(x-1)\delta^\times(x)=0$.
Equation~\eqref{eq:twopoint} defines a formal power series in $u$ whose components belong to $\mathbb{C}[z^{\pm1},\bz^{\pm1}][[u]]$.
At each order in $u$, the exponential is a polynomial in $t_1,t_2$, so each component is well defined.
Contact terms supported at $z\bz=0$ are excluded.

The $P_{--}$ Ward identity implies that $G$ depends on $u_1,u_2$ only through their difference $u=u_1-u_2$.
The Ward identities for the translations generated by $P_{+-}$ and $P_{-+}$ read
\begin{equation}
    \bigl(z\partial_u+\xi(t_1+t_2)\bigr)G=0
    \, ,\qquad
    \bigl(\bz\partial_u+\xi(t_1^{-1}+t_2^{-1})\bigr)G=0
    \, .
    \label{eq:wardtranslation}
\end{equation}
Subtracting $\bz$ times the first equation from the $P_{++}$ Ward identity and multiplying by $t_1$ gives $(z-\bz t_1t_2)G=0$.
At $u=0$, the Ward identities for the Cartan and lowering generators require the two component indices to coincide and determine the powers of $z$ and $\bz$.
The relation $(z-\bz t_1t_2)G=0$ fixes the relative normalizations of these components.
Extracting them by residues gives
\begin{equation}
    \operatorname{Res}_{t_1=0}
    \operatorname{Res}_{t_2=0}
    t_1^{-n}t_2^{-m}\left.G\right|_{u=0}
    \frac{\dd t_1}{t_1}\frac{\dd t_2}{t_2}
    =C\delta_{nm}z^{-1-n}\bz^{-1+n}
    \, .
    \label{eq:twoinitial}
\end{equation}
Each residue selects the corresponding Laurent coefficient, as defined in Appendix~\ref{app:deltaformal}.
The first equation in \eqref{eq:wardtranslation} determines the coefficients at higher orders in $u$, yielding \eqref{eq:twopoint} uniquely in the stated class, up to $C$.
Direct substitution verifies the remaining Lorentz Ward identities and the null-state equation at both insertions, using $(1-x)(\delta^\times)'(x)=\delta^\times(x)$.
Appendix~\ref{app:deltaformal} gives a coefficientwise proof of the delta identities and the null-state equation.

The two-point function also realizes the BPZ form introduced above.
Applying \eqref{eq:opeBPZinversion} to its second insertion and taking the primary-state coefficients gives $B_0\delta^\times(t_1t_2)$, after matching $C$ to $B_0$.
Comparing the coefficients of $t_1^nt_2^m$ therefore yields $\hat{B}([e_n],[e_m])=B_0\delta_{nm}$, in agreement with \eqref{eq:bpz}.
This fixes the normalization of the outgoing dual.

\subsection{Three-point functions}
\label{subsec:threepoint}
We take three generating operators $\OO_i^{\rm b}(u_i,z_i,\bz_i,t_i)$ in $\SSS_{0,\xi_i}, i=1,2,3$, with independent $\xi_i\ne0$, and write their three-point function as $G_3=\langle\OO_1^{\rm b}\OO_2^{\rm b}\OO_3^{\rm b}\rangle$.
The action at the $i$-th insertion is \eqref{eq:localcartan}--\eqref{eq:locallowering} with $\xi$ replaced by $\xi_i$.
Symmetry and the null-state equations require
\begin{equation}
    \sum_{i=1}^3\D_X^{\rm b,(i)}G_3=0
    \, ,\qquad
    Q_{\xi_i}^{(i)}G_3=0\quad(i=1,2,3)
    \, .
    \label{eq:threepointward}
\end{equation}
We work with separated insertions, $z_{ij}\bz_{ij}\ne0$, on the chart $D_{123}=z_{13}\bz_{23}-z_{23}\bz_{13}\ne0$, where $z_{ij}=z_i-z_j$ and similarly for $u_{ij},\bz_{ij}$.
Geometrically, $D_{123}$ tests the linear independence of the two spatial displacements from point $3$; on the real slice $\bz=z^*$, it is proportional to the oriented area of the triangle.
It is also the determinant of the coefficient matrix of the two translation Ward equations below, so $D_{123}\ne0$ allows us to invert that matrix and determine the dependence on $u_{13},u_{23}$.
We restrict to this nondegenerate case; the locus $D_{123}=0$ requires a separate analysis.

The Ward identities associated with $L_-$, $\bL_-$ and $P_{--}$ reduce the spacetime dependence to differences relative to point $3$.
The Ward identities for the translations generated by $P_{+-}$ and $P_{-+}$ read
\begin{equation*}
    \begin{aligned}
    \left(z_{13}\partial_{u_{13}}+z_{23}\partial_{u_{23}}
    +\sum_{i=1}^3\xi_it_i\right)G_3&=0
    \, ,\\
    \left(\bz_{13}\partial_{u_{13}}+\bz_{23}\partial_{u_{23}}
    +\sum_{i=1}^3\xi_it_i^{-1}\right)G_3&=0
    \, .
    \end{aligned}
\end{equation*}
Inverting this two-by-two system determines the dependence of $G_3$ on $u_{13},u_{23}$:
\begin{equation}
    \begin{aligned}
    G_3&=e^{\mathcal{E}_{123}}H
    \, ,\\
    \mathcal{E}_{123}
    &=\frac{1}{D_{123}}\sum_{i=1}^3\xi_i
    \left[u_{13}(z_{23}t_i^{-1}-\bz_{23}t_i)
    +u_{23}(\bz_{13}t_i-z_{13}t_i^{-1})\right]
    \, ,
    \end{aligned}
    \label{eq:threepointexponent}
\end{equation}
where $H$ is independent of $u_{13},u_{23}$.
Substitution into the Ward identity associated with $P_{++}$ gives $\mathcal{C}_{123}H=0$, with
\begin{equation}
    \mathcal{C}_{123}
    =\det
    \begin{pmatrix}
        1&1&1&0\\
        z_1&z_2&z_3&\displaystyle\sum_{i=1}^3\xi_it_i\\[4pt]
        \bz_1&\bz_2&\bz_3&\displaystyle\sum_{i=1}^3\xi_it_i^{-1}\\[4pt]
        z_1\bz_1&z_2\bz_2&z_3\bz_3&
        \displaystyle\sum_{i=1}^3\xi_i(\bz_it_i+z_it_i^{-1})
    \end{pmatrix}
    \, .
    \label{eq:threepointcarrier}
\end{equation}
At fixed insertion points and parameters $\xi_i$, this is a Laurent polynomial in the nonzero complex auxiliary variables $t_i$.
Define
\begin{equation}
    \Sigma=\left\{(t_1,t_2,t_3)\in(\mathbb{C}^\times)^3:
    \mathcal{C}_{123}(t_1,t_2,t_3)=0\right\}
    \, .
    \label{eq:threepointsupport}
\end{equation}
Wherever $\mathcal{C}_{123}\ne0$, its reciprocal is locally holomorphic, and $\mathcal{C}_{123}H=0$ forces $H$ to vanish there.
Thus the support of $H$, and hence of $G_3$, must lie in $\Sigma$.
This is a constraint in the auxiliary variables $t_i$; the spacetime insertion points enter as parameters.
The distribution on this locus requires an additional prescription, which we now specify.

We work on the part of $\Sigma$ where
\begin{equation*}
    \dd_t\mathcal{C}_{123}
    =\sum_{i=1}^3\frac{\partial\mathcal{C}_{123}}{\partial t_i}\,\dd t_i
    \ne0
    \, .
\end{equation*}
This means that at least one partial derivative is nonzero at the point under consideration, even though $\mathcal{C}_{123}=0$ there.
For example, if $\partial_{t_3}\mathcal{C}_{123}\ne0$, the constraint locally determines $t_3$ as a holomorphic function of $t_1,t_2$.
The surface then has two independent complex coordinates, and $\mathcal{C}_{123}$ can serve as a coordinate transverse to it.
The nonzero differential is a regularity assumption for the residue construction, rather than an additional Ward equation.

On this regular part of $\Sigma$, we choose a holomorphic residue realization with the logarithmic volume form $\Omega_t=(\dd t_1/t_1)\wedge(\dd t_2/t_2)\wedge(\dd t_3/t_3)$.
The residue uses a small positively oriented contour around zero in the transverse coordinate, holding the two coordinates along $\Sigma$ fixed.
A solution of \eqref{eq:threepointward} is
\begin{equation}
    G_3=C_3(\xi_1,\xi_2,\xi_3)\,
    e^{\mathcal{E}_{123}}\delta_{\mathbb{C}}(\mathcal{C}_{123})
    \, .
    \label{eq:threepointsolution}
\end{equation}
Here $\delta_{\mathbb{C}}(\mathcal{C}_{123})$ denotes the hypersurface residue distribution defined relative to $\Omega_t$ in Appendix~\ref{app:deltaleray}; it evaluates only the transverse coordinate and leaves a two-form on $\Sigma$.
Multiplication by $\mathcal{C}_{123}$ removes the pole and gives zero residue, implementing $\mathcal{C}_{123}H=0$.
The normalization of $\delta_{\mathbb{C}}$ is fixed by this residue convention, while $C_3$ is an undetermined three-point coefficient.

Direct substitution verifies the Ward identities associated with $L_0,\bL_0,L_+,\bL_+$ using $\mathcal{C}_{123}\delta'_{\mathbb{C}}(\mathcal{C}_{123})=-\delta_{\mathbb{C}}(\mathcal{C}_{123})$ and $\mathcal{C}_{123}\delta_{\mathbb{C}}(\mathcal{C}_{123})=0$ from Appendix~\ref{app:deltaleray}.
The null-state equation at each insertion reduces to
\begin{equation}
    \begin{aligned}
    Q_{\xi_i}^{(i)}\bigl[e^{\mathcal{E}_{123}}
    \delta_{\mathbb{C}}(\mathcal{C}_{123})\bigr]
    &=\frac{\xi_i e^{\mathcal{E}_{123}}}{D_{123}}
    \left(\partial_{\bz_i}D_{123}-t_i^2\partial_{z_i}D_{123}\right)\\
    &\quad\times\left[
        \delta_{\mathbb{C}}(\mathcal{C}_{123})
        +\mathcal{C}_{123}\delta'_{\mathbb{C}}(\mathcal{C}_{123})
    \right]=0
    \, .
    \end{aligned}
    \label{eq:threepointnullcheck}
\end{equation}
This identity follows by differentiating \eqref{eq:threepointexponent}--\eqref{eq:threepointcarrier} and using \eqref{eq:deltalerayidentities}; derivatives at point $3$ act on the differences with the opposite sign.
This local solution holds on the smooth support for arbitrary nonzero $\xi_i$.

Appendix~\ref{app:ope} uses this three-point solution to compute the primary and descendant coefficients in the formal OPE.

\section{The intertwiner for the bulk massive scalar}
\label{sec:bulk}

\subsection{The bulk representation and intertwining equations}
We now seek an explicit integral transform from bulk momentum-space operators to the local Carrollian fields constructed above.
Requiring the map to preserve the action of every \Poincare generator gives \eqref{eq:targetintertwining}.
Once both differential actions are specified, this requirement becomes a system of equations for the kernel.
We first write the bulk action on the mass shell.

We parametrize a future-directed momentum of mass $m>0$ using celestial coordinates $(y',z',\bz')$, which are upper-half-space coordinates for $p^\mu/m$ on the unit future hyperboloid:
\begin{equation}
    p^\mu(y',z',\bz')=\frac{m}{2y'}
    \bigl(1+y'^2+z'\bz',\ z'+\bz',\ -\ii(z'-\bz'),\
    1-y'^2-z'\bz'\bigr)
    \, ,
    \label{eq:massshell}
\end{equation}
with $y'>0$ and $p^2=-m^2$.
On the real mass shell, $\bz'=(z')^*$.
These primed variables label bulk momenta and are distinct from the boundary spacetime coordinates $(u,z,\bz)$.
To construct the intertwiner, we locally complexify these coordinates and treat $z'$ and $\bz'$ as independent variables.
We use the measure
\begin{equation}
    \dd\mu_m=\frac{m^2}{2}y'^{-3}\dd y'\dd z'\dd\bz'
    \, ,
    \label{eq:measure}
\end{equation}
where the constant associated with the complex-coordinate orientation can be absorbed into the normalization of the transform.
Set the Fourier normalization to one and write $s=+1$ for outgoing and $s=-1$ for incoming operators.

The bulk Lorentz action on the scalar momentum-space operator $\OO^{\rm B}$ is
\begin{align}
    \D^{\rm B}_{L_n}
    &=z'^{n+1}\partial_{z'}+\frac{n+1}{2}z'^ny'\partial_{y'}
    -\frac{n(n+1)}{2}z'^{n-1}y'^2\partial_{\bz'},\notag\\
    \D^{\rm B}_{\bL_n}
    &=\bz'^{n+1}\partial_{\bz'}+\frac{n+1}{2}\bz'^ny'\partial_{y'}
    -\frac{n(n+1)}{2}\bz'^{n-1}y'^2\partial_{z'}
    \, ,
    \label{eq:bulkLorentz}
\end{align}
with terms of zero coefficient omitted.
The translations multiply by
\begin{equation}
    \bigl(\D^{\rm B}_{P_{--}},\D^{\rm B}_{P_{+-}},
    \D^{\rm B}_{P_{-+}},\D^{\rm B}_{P_{++}}\bigr)
    =-\frac{\ii s m}{y'}\bigl(1,z',\bz',z'\bz'+y'^2\bigr)
    \, .
    \label{eq:bulktranslation}
\end{equation}
All six vector fields in \eqref{eq:bulkLorentz} have zero divergence with respect to \eqref{eq:measure}.
For example, $\partial_{z'}(z'y'^{-3})+ \partial_{y'}(\tfrac{1}{2} y'^{-2})=0$ for $L_0$.

We now construct an integral transform that satisfies the intertwining equation \eqref{eq:targetintertwining}, using the boundary differential action \eqref{eq:localcartan}--\eqref{eq:locallowering}:
\begin{equation}
    \OO^{\rm b}(u,z,\bz,t)
    =\int\dd\mu_m\,K(u,z,\bz,t;y',z',\bz')\OO^{\rm B}(y',z',\bz')
    \, .
    \label{eq:integralmap}
\end{equation}
Here $K(u,z,\bz,t;y',z',\bz')$ is the integral kernel of the linear map $K$.
In the following kernel equations, $\D_X^{\rm b}$ acts on the boundary coordinates and $\D_X^{\rm B}$ on the bulk momentum coordinates.
Covariance requires $\D_X^{\rm b} K=(\D_X^{\rm B})^{\mathsf{T}}K$, where $\mathsf{T}$ is the formal transpose for this measure.
Hence the ten intertwining equations are
\begin{equation}
    (\D_L^{\rm b}+\D_L^{\rm B})K=0\quad(L=L_n,\bL_n)
    \, ,\qquad
    (\D_P^{\rm b}-\D_P^{\rm B})K=0\quad(P=P_{k,l})
    \, .
    \label{eq:intertwinersystem}
\end{equation}
Integration by parts requires all boundary terms to vanish.
The $z'$ and $\bz'$ integrations may use local complex contours; smooth radial test functions with compact support in $(0,\infty)$ eliminate the radial endpoint terms.
By Section~\ref{sec:targetrequirements}, a solution of all the equations maps the bulk scalar into fields annihilated by every $\D_{W^\mu}^{\rm b}$.
The quotient analysis and \eqref{eq:PLnullbridge} then show that these fields obey the null-state equation associated with $\SSS_{0,\xi}$.
We therefore solve the intertwining equations without assuming the null-state equation as an additional constraint on the kernel.

\subsection{Constructing the kernel}
\label{subsec:solve}

The $P_{--}$ equation fixes the dependence of $K$ on $u$ through the factor $e^{-\ii smu/y'}$.
The $P_{+-}$ and $P_{-+}$ equations impose $AK=\bar{A} K=0$, where
\begin{equation}
    A=z'-z-\frac{\ii s\xi}{m}y't
    \, ,\qquad
    \bar{A}=\bz'-\bz-\frac{\ii s\xi}{m}y't^{-1}
    \, .
    \label{eq:characteristic}
\end{equation}
For a kernel represented by evaluation functionals on locally holomorphic functions, these constraints give
\begin{equation}
    K=f(y',t)e^{-\ii smu/y'}\delta_{\mathbb{C}}(A)
    \delta_{\mathbb{C}}(\bar{A})
    \, .
    \label{eq:kernelansatz}
\end{equation}
The $L_-$ and $\bL_-$ equations make the prefactor $f$ independent of $z$ and $\bz$.
We define $\delta_{\mathbb{C}}(w-a):\ \phi\longmapsto\phi(a)$ for locally holomorphic $\phi$, with $w\delta'_{\mathbb{C}}(w)=-\delta_{\mathbb{C}}(w)$.
Thus $\delta_{\mathbb{C}}$ denotes holomorphic evaluation, which may take place at complex points off the real mass shell.
We take $f$ to be regular on a connected domain with $y'>0$ and $t\in\mathbb{C}^\times$.

For $L_0$, the combined vector field in \eqref{eq:intertwinersystem} assigns weight $1$ to $A$ and weight $0$ to $\bar{A}$, and annihilates the exponential factor.
Including the boundary weight $1/2$ gives the first equation below; the $\bL_0$ equation gives the second:
\begin{equation}
    (y'\partial_{y'}+t\partial_t-1)f=0
    \, ,\qquad
    (y'\partial_{y'}-t\partial_t-1)f=0
    \, .
    \label{eq:density}
\end{equation}
It follows that $f=N_{m,\xi,s}y'$, with $N_{m,\xi,s}$ independent of all coordinates.
Substituting this result into the remaining translation equation gives
\begin{equation}
    (\D_{P_{++}}^{\rm b}-\D^{\rm B}_{P_{++}})K
    =\ii s m y'\left(1-\frac{\xi^2}{m^2}\right)K
    \, .
    \label{eq:massobstruction}
\end{equation}
Since $y'>0$, a nonzero kernel requires
\begin{equation}
    \xi^2=m^2
    \, .
    \label{eq:massmatching}
\end{equation}
This is also required by equality of the mass Casimirs: \eqref{eq:bulktranslation} gives $C_{\rm mass}^{\rm B}=m^2$, whereas \eqref{eq:masscasimir} gives $\xi^2$.
Notice that the positive eigenvalue is consistent with $p^2=-m^2$ because the translation generators carry the factor $\ii$.

Writing $\xi=\sigma m$, $\sigma=\pm1$, these equations give the candidate kernel
\begin{equation}
    K_{\sigma m}^{s,m}
    =N_{m,\sigma,s}y'e^{-\ii smu/y'}
    \delta_{\mathbb{C}}(z'-z-\ii s\sigma y't)\times
    \delta_{\mathbb{C}}(\bz'-\bz-\ii s\sigma y't^{-1})
    \, .
    \label{eq:kernel}
\end{equation}
The shifts in $z'$ and $\bz'$ have the same sign.
Applying the evaluation functionals gives the integral transform
\begin{equation}
    \OO^{\rm b}(u,z,\bz,t)
    =\frac{N_{m,\sigma,s}m^2}{2}
    \int_0^\infty\frac{\dd y'}{y'^2}e^{-\ii smu/y'}
    \OO^{\rm B}(y',z+\ii s\sigma y't,\bz+\ii s\sigma y't^{-1})
    \, .
    \label{eq:evaluationmap}
\end{equation}

The normalization $N_{m,\sigma,s}$ remains free for each chosen branch.
The two signs of $\xi$ are related algebraically by $e_n\mapsto(-1)^ne_n$ (or $t\mapsto-t$), whereas $s$ records the incoming/outgoing bulk convention.
The complex shifts in \eqref{eq:kernel} generally take the momentum coordinates away from the real mass shell.

To complete the kernel construction, let $K_\xi$ denote the candidate integral transform with \eqref{eq:kernelansatz} and $f=N_{m,\xi,s}y'$, temporarily retaining arbitrary $\xi\ne0$.
Acting on $\OO^{\rm B}$, the remaining $L_1$ and $\bL_1$ equations reduce, after radial integration by parts, to
\begin{align}
    (\D_{L_1}^{\rm b}K_\xi-K_\xi\D^{\rm B}_{L_1})\OO^{\rm B}
    &=\left(1-\frac{\xi^2}{m^2}\right)
    K_\xi\!\left(y'^2\partial_{\bz'}\OO^{\rm B}\right),\notag\\
    (\D_{\bL_1}^{\rm b}K_\xi-K_\xi\D^{\rm B}_{\bL_1})\OO^{\rm B}
    &=\left(1-\frac{\xi^2}{m^2}\right)
    K_\xi\!\left(y'^2\partial_{z'}\OO^{\rm B}\right)
    \, .
    \label{eq:loweringresidual}
\end{align}
The boundary terms vanish on the domain specified above.
Both residuals vanish for $\xi^2=m^2$, completing all the intertwining equations.

\section{Discussion}
\label{sec:discussion}

We have studied how a four-dimensional bulk massive scalar can be represented by three-dimensional Carrollian fields.
\Poincare covariance requires the image of a holographic map to inherit both the bulk Casimir eigenvalues and the vanishing Pauli--Lubanski vector.
The latter requirement is essential: in the family of Carrollian modules considered here, matching the Casimirs alone permits $\Delta=1,2$, whereas imposing the Pauli--Lubanski condition selects $\Delta=1$.
At this weight, the relations required by $W^\mu=0$ generate precisely the radical of the BPZ form.
Removing these null states therefore yields an irreducible quotient representation with the required mass and spin properties.

The usual construction from a Carrollian primary multiplet and its descendants gives the module $\MM_{1/2,1/2,\xi}$ in the scalar case considered here. An open question is how to construct a Carrollian field theory in which the null-state equation $Q_\xi\OO^{\rm b}=0$ follows from the field equations, so that the scalar operators realize the quotient $\SSS_{0,\xi}=\MM_{1/2,1/2,\xi}/\mathcal{N}$.

For the symmetric weights $h=\bar{h}=1/2$, the state--operator correspondence turns these relations into the local equation $Q_\xi\OO^{\rm b}=0$.
We have obtained a formal two-point function and a three-point solution of the Ward and null-state equations, together with the associated formal OPE contributions.
On the specified domain of analytically continued bulk momentum-space data, the integral kernel satisfies all the intertwining equations for $\xi^2=m^2$ and produces boundary fields obeying the same local relations.

The integral transform evaluates the bulk field at shifted momenta. Whether these momenta can be real depends on the spacetime signature.
The evaluation conditions in \eqref{eq:kernel} imply
\begin{equation}
    (z'-z)(\bz'-\bz)=-y'^2
    \, ,\qquad y'>0\, .
    \label{eq:relativesupport}
\end{equation}
On the Lorentzian real slice, $\bz=z^*$ and $\bz'=(z')^*$, the left-hand side is $|z'-z|^2\ge0$.
Hence this kernel has no real support on that slice and requires analytic continuation of the momentum-space data.
At the boundary origin, \eqref{eq:relativesupport} reduces to $z'\bz'=-y'^2$; at a general insertion point it constrains the relative coordinates $z'-z$ and $\bz'-\bz$.
The same support condition admits real solutions in Klein signature $(2,2)$, where $z,\bz$ and $z',\bz'$ are independent real coordinates~\cite{Atanasov2021,ChenHuMao2026}.
In our conventions, the auxiliary variable must also satisfy $t=\ii\tau$ with $\tau\in\mathbb{R}\setminus\{0\}$.
The two evaluation conditions then become
\begin{equation}
    z'=z-s\sigma y'\tau
    \, ,\qquad
    \bz'=\bz+\frac{s\sigma y'}{\tau}
    \, ,
    \label{eq:kleinsupport}
\end{equation}
which have real solutions and obey \eqref{eq:relativesupport}.
Keeping $t$ real would instead leave both shifts imaginary.
The bulk real structure changes simultaneously: setting $p_{\rm K}^{2}=\ii p_{\rm L}^{2}$ and leaving the other Cartesian components unchanged converts \eqref{eq:massshell} into a real momentum with metric $\eta_{\rm K}=\operatorname{diag}(-1,1,-1,1)$ and $\eta^{\rm K}_{\mu\nu}p_{\rm K}^{\mu}p_{\rm K}^{\nu}=-m^2$.
Thus the Klein real slice admits real support for this kernel.
An integral representation using ordinary real delta functions still requires a compatible integration domain, measure and normalization.

In~\cite{Liu2026}, Carrollian amplitudes were defined by integrating a bulk scattering amplitude against the Carrollian basis of each external particle.
The kernel constructed here plays the role of this basis for a four-dimensional massive scalar.
Because it evaluates bulk data at complex momenta, defining massive Carrollian amplitudes in this way requires the integration prescription discussed above.
A first test would be to apply the transform to each external leg of a three-point amplitude for a cubic interaction of massive scalars. One could compare an integration over real Klein momenta with a contour prescription on the complexified mass shell. For each prescription, does the transformed amplitude reproduce the support $\mathcal{C}_{123}=0$ and the dependence on Carrollian time in \eqref{eq:threepointsolution}? If so, the comparison would relate the coefficient $C_3$ to the bulk cubic coupling once the external fields are normalized.

The map constructed here acts from the bulk to the boundary, $\OO^{\rm b}=K\OO^{\rm B}$.
If $K$ is injective on the chosen bulk domain, it admits an inverse on its image, so the resulting boundary field uniquely determines the bulk data:
\begin{equation*}
    \OO^{\rm B}=K^{-1}\OO^{\rm b}
    \, ,\qquad
    \OO^{\rm b}\in\operatorname{Im}K
    \, .
\end{equation*}
The inverse $K^{-1}$ would provide bulk reconstruction in momentum space.
If $K$ is also surjective onto the chosen boundary solution space, this reconstruction would apply to every field in that space.
Injectivity and surjectivity require input beyond \Poincare covariance.
Determining these properties and constructing $K^{-1}$ are natural next steps.

Another direction is to extend the bulk-to-boundary transform to massive particles with spin. At fixed momentum, the Pauli--Lubanski vector acts on the spin indices of the massive one-particle states. The boundary representation must reproduce this action.

\begin{acknowledgments}
    The work is partly supported by NSFC Grant No. 12588101, 12405082, and No. 12275004.
    WJM is also supported by  Shanghai Pujiang Program No. 24PJA118.
    This work is financially supported by the Ningbo Science and Technology Bureau under program grant No. 2024DJRC010009.
    We use GPT-5.6 Sol and GPT-6 Astra to carry out certain derivations and to help draft the manuscript.
\end{acknowledgments}

\appendix
\section{Delta distributions}
\label{app:formaldelta}

\subsection{The formal delta \texorpdfstring{$\delta^\times$}{delta-times}}
\label{app:deltaformal}

The bilateral series $\delta^\times(x)=\sum_{r\in\mathbb{Z}}x^r$ in \eqref{eq:twopoint} is understood coefficientwise, without asserting pointwise convergence.
Here $\operatorname{Res}$ denotes the formal residue, which extracts the coefficient of $t^{-1}\dd t$:
\begin{equation*}
    \operatorname{Res}_{t=0}t^k\dd t
    =\delta_{k,-1}
    \, ,\qquad k\in\mathbb{Z}\, .
\end{equation*}
In \eqref{eq:twoinitial}, the two residues select the terms with $r=n$ and $r=m$ in the bilateral series, giving $\delta_{nm}$.
Multiplication by an integer power only shifts the summation index:
\begin{equation*}
    x^m\delta^\times(x)
    =\sum_{r\in\mathbb{Z}}x^{r+m}
    =\sum_{k\in\mathbb{Z}}x^k
    =\delta^\times(x)
    \, ,\qquad m\in\mathbb{Z}\, .
\end{equation*}
A Laurent polynomial is a finite sum of such powers, so linearity gives the first identity below.
Taking $\phi(x)=1-x$ gives the second:
\begin{equation}
    \phi(x)\delta^\times(x)=\phi(1)\delta^\times(x)
    \, ,\qquad
    (1-x)\delta^\times(x)=0
    \, ,\qquad
    (1-x)(\delta^\times)'(x)=\delta^\times(x)
    \, .
    \label{eq:deltaformalidentities}
\end{equation}
The third follows by differentiating the second coefficientwise:
\begin{equation*}
    0=\frac{\dd}{\dd x}\bigl[(1-x)\delta^\times(x)\bigr]
    =-\delta^\times(x)+(1-x)(\delta^\times)'(x)
    \, .
\end{equation*}

These rules apply order by order in $u$ to \eqref{eq:twopoint}, since each coefficient of the exponential is a polynomial in $t_1,t_2$.
In particular, setting $x=\bz t_1t_2/z$, the two null-state equations reduce to
\begin{equation}
    Q_\xi^{(1)}G=-Q_\xi^{(2)}G
    =\frac{\xi C}{z\bz^2}e^{-\xi u(t_1+t_2)/z}
    \left[x(1-x)(\delta^\times)'(x)-\delta^\times(x)\right]=0
    \, .
    \label{eq:twopointdeltanullcheck}
\end{equation}
Here $u,z,\bz$ are the coordinate differences used in Section~\ref{subsec:twopoint}.
The bracket becomes $x\delta^\times(x)-\delta^\times(x)=(x-1)\delta^\times(x)=0$, using both identities in \eqref{eq:deltaformalidentities}.

\subsection{Holomorphic evaluation}
\label{app:deltaholomorphic}

The kernel uses $\delta_{\mathbb{C}}$ to evaluate holomorphic test functions at a possibly complex point:
\begin{equation}
    \delta_{\mathbb{C}}(w-a):\ \phi
    \longmapsto\phi(a)=\frac{1}{2\pi\ii}\oint_{\Gamma_a}
    \frac{\phi(w)}{w-a}\,\dd w
    \, .
    \label{eq:deltaholomorphicevaluation}
\end{equation}
Here $\phi$ is holomorphic inside a small positively oriented contour $\Gamma_a$ around $a$.
To see why this gives evaluation, expand locally as $\phi(w)=\phi(a)+\phi'(a)(w-a)+O((w-a)^2)$.
Only the constant term contributes to the contour integral in \eqref{eq:deltaholomorphicevaluation}; the other terms have zero residue.
The prime denotes the distributional derivative in the argument $w-a$, so its action is
\begin{equation*}
    \delta'_{\mathbb{C}}(w-a):\ \phi
    \longmapsto-\phi'(a)
    =-\frac{1}{2\pi\ii}\oint_{\Gamma_a}
    \frac{\phi(w)}{(w-a)^2}\,\dd w
    \, .
\end{equation*}
The double pole now selects the linear term of the same expansion.

Multiplying the test function by $w-a$ gives
\begin{equation*}
    \begin{aligned}
    (w-a)\delta_{\mathbb{C}}(w-a):\ \phi&\longmapsto0
    \, ,\\
    (w-a)\delta'_{\mathbb{C}}(w-a):\ \phi
    &\longmapsto-\left.\frac{\dd}{\dd w}\bigl[(w-a)\phi(w)\bigr]\right|_{w=a}
    =-\phi(a)
    \, .
    \end{aligned}
\end{equation*}
Since these hold for every holomorphic test function, they establish
\begin{equation}
    (w-a)\delta_{\mathbb{C}}(w-a)=0
    \, ,\qquad
    (w-a)\delta'_{\mathbb{C}}(w-a)=-\delta_{\mathbb{C}}(w-a)
    \, .
    \label{eq:deltaholomorphicidentities}
\end{equation}
The two factors in \eqref{eq:kernel} apply this evaluation separately to $z'$ and $\bz'$.
This is a holomorphic functional, not an ordinary Dirac distribution at a nonreal point on the real mass shell.

\subsection{The three-point constraint}
\label{app:deltaleray}

The three-point support lies in the surface $\Sigma$ defined in \eqref{eq:threepointsupport}.
Fix the insertion points and parameters $\xi_i$, with $D_{123}\ne0$, and regard $t_1,t_2,t_3$ as independent variables.
Consider a point of $\Sigma$ where $\dd_t\mathcal{C}_{123}\ne0$.
The constraint function itself can be used as a transverse coordinate.
For instance, on a patch where $\partial_{t_3}\mathcal{C}_{123}\ne0$, the change of variables
\begin{equation*}
    (t_1,t_2,t_3)\longmapsto(t_1,t_2,\mathcal{C}_{123})
\end{equation*}
has nonzero Jacobian $\partial_{t_3}\mathcal{C}_{123}$.
Thus $t_1,t_2,\mathcal{C}_{123}$ are independent local coordinates, and $t_3$ is locally a holomorphic function of all three.
In these coordinates $\Sigma$ is simply $\mathcal{C}_{123}=0$, or equivalently $t_3=g(t_1,t_2)$ on the surface.
The term normal coordinate refers here to the transverse variable $\mathcal{C}_{123}$; no metric or notion of orthogonality is required.
The differential is computed before restricting to $\Sigma$: it is nonzero in the original three-variable space, while it vanishes on variations tangent to $\Sigma$.

With $\Omega_t=\bigwedge_{i=1}^3\dd t_i/t_i$, the holomorphic evaluation of Appendix~\ref{app:deltaholomorphic} in the normal coordinate gives the local residue realization
\begin{equation}
    \begin{aligned}
    \delta_{\mathbb{C}}(\mathcal{C}_{123})\Omega_t:\ \phi
    &\longmapsto\operatorname{Res}_{\mathcal{C}_{123}=0}
    \frac{\phi\Omega_t}{\mathcal{C}_{123}}
    \, ,\\
    \delta'_{\mathbb{C}}(\mathcal{C}_{123})\Omega_t:\ \phi
    &\longmapsto-\operatorname{Res}_{\mathcal{C}_{123}=0}
    \frac{\phi\Omega_t}{\mathcal{C}_{123}^{\,2}}
    \, .
    \end{aligned}
    \label{eq:deltaleraydefinition}
\end{equation}
The residue is taken in $\mathcal{C}_{123}$, holding $t_1,t_2$ fixed; it extracts the coefficient of $\dd\mathcal{C}_{123}/\mathcal{C}_{123}$ and leaves a two-form along the support.
The double-pole result is a representative in this chosen normal coordinate chart.

To evaluate these residues explicitly, write the test top form locally as $\phi\Omega_t=\dd\mathcal{C}_{123}\wedge\omega(t_1,t_2;\mathcal{C}_{123})$.
Here $\omega$ is a two-form in $t_1,t_2$ whose coefficient is holomorphic in the three adapted coordinates and includes both the test function and the measure density.
Then
\begin{equation*}
    \begin{aligned}
    \operatorname{Res}_{\mathcal{C}_{123}=0}
    \frac{\dd\mathcal{C}_{123}\wedge\omega}
    {\mathcal{C}_{123}}
    &=\omega(t_1,t_2;0)
    \, ,\\
    -\operatorname{Res}_{\mathcal{C}_{123}=0}
    \frac{\dd\mathcal{C}_{123}\wedge\omega}
    {\mathcal{C}_{123}^{\,2}}
    &=-\left.\partial_{\mathcal{C}_{123}}\omega\right|_{\mathcal{C}_{123}=0}
    \, .
    \end{aligned}
\end{equation*}
The first equality is the Cauchy evaluation
\begin{equation*}
    \frac{1}{2\pi\ii}\oint_{|\mathcal{C}_{123}|=\epsilon}
    \frac{\omega(t_1,t_2;\mathcal{C}_{123})}{\mathcal{C}_{123}}
    \,\dd\mathcal{C}_{123}=\omega(t_1,t_2;0)
    \, .
\end{equation*}
The small contour is positively oriented.
Both the contour integral and $\partial_{\mathcal{C}_{123}}\omega$ hold $t_1,t_2$ fixed; the derivative acts on the complete coefficient of $\omega$, including the measure density.

The Jacobian can be displayed explicitly in the patch above:
\begin{equation*}
    \phi\Omega_t
    =\dd\mathcal{C}_{123}\wedge
    \frac{\phi\,\dd t_1\wedge\dd t_2}
    {t_1t_2t_3\,\partial_{t_3}\mathcal{C}_{123}}
    \, .
\end{equation*}
Taking the residue therefore gives
\begin{equation}
    \operatorname{Res}_{\mathcal{C}_{123}=0}
    \frac{\phi\Omega_t}{\mathcal{C}_{123}}
    =\left.
    \frac{\phi}{t_1t_2t_3\,\partial_{t_3}\mathcal{C}_{123}}
    \right|_{t_3=g(t_1,t_2)}\dd t_1\wedge\dd t_2
    \, .
    \label{eq:deltalerayjacobian}
\end{equation}
The residue thus fixes one complex variable and leaves a two-form in the other two.
The nonzero partial derivative is precisely what makes this local coordinate change and its Jacobian well defined.
If another partial derivative is nonzero, that variable can be used instead.
Points where all three vanish are outside this prescription; their exclusion is not a consequence of the Ward equation.

Replacing $\phi$ by $\mathcal{C}_{123}\phi$ replaces $\omega(t_1,t_2;\mathcal{C}_{123})$ by $\mathcal{C}_{123}\omega(t_1,t_2;\mathcal{C}_{123})$, giving zero in the first residue and $-\omega(t_1,t_2;0)$ in the second.
This proves
\begin{equation}
    \mathcal{C}_{123}\delta_{\mathbb{C}}(\mathcal{C}_{123})=0
    \, ,\qquad
    \mathcal{C}_{123}\delta'_{\mathbb{C}}(\mathcal{C}_{123})
    =-\delta_{\mathbb{C}}(\mathcal{C}_{123})
    \, .
    \label{eq:deltalerayidentities}
\end{equation}
These are the identities used in the three-point Ward and null-state checks.
The Jacobian in \eqref{eq:deltalerayjacobian} has no absolute value because this is a holomorphic contour residue.

The residue in \eqref{eq:deltalerayjacobian} is a two-form on $\Sigma$, rather than a number.
To obtain a scalar pairing with a holomorphic test function, one may further specify a closed oriented real two-cycle $\gamma$ in the regular part of $\Sigma$ and set
\begin{equation*}
    \phi\longmapsto
    \int_\gamma\operatorname{Res}_{\mathcal{C}_{123}=0}
    \frac{\phi\Omega_t}{\mathcal{C}_{123}}
    \, .
\end{equation*}
The integrand must be defined along the cycle and the integral must exist.
On a patch with $t_3=g(t_1,t_2)$, such a cycle can be obtained by lifting a two-cycle in the remaining variables $t_1,t_2$, provided the chosen branch closes along it.
This integration along $\Sigma$ is distinct from the small transverse contour used to take the residue.
Homologous cycles give the same pairing when the residue form is single-valued and holomorphic throughout the deformation region; different cycle classes can give different pairings.
The Ward equations and the condition $\dd_t\mathcal{C}_{123}\ne0$ do not select $\gamma$.
A global prescription requires additional analytic or physical input, including compatible transport as the insertion points vary.

Appendix~\ref{app:ope} instead keeps $t_1,t_2$ as external auxiliary variables and takes only a residue in the third variable, around one selected simple root.
After the BPZ inversion, the pairing with $\delta^\times(t/t_3)$ in \eqref{eq:opeTauInvertedPairing} leaves the logarithmic two-form with $\tau_{12}(\lvert\OO_3(t)\rangle;1,2)$ as its coefficient, understood coefficientwise in $t$.
No integration over $\gamma$ is needed for this partial pairing.
The choice of the third root and the subsequent choice of a cycle for the remaining variables are therefore separate data.

\section{Formal OPE and descendant coefficients}
\label{app:ope}

\subsection{Three-point pairings and descendant data}
\label{app:opeprimary}

We reconstruct the contribution of $\SSS_{0,\xi_3}$, carried by the third field $\OO_3^{\rm b}$ in Section~\ref{subsec:threepoint}, to the OPE of $\OO_1^{\rm b}$ and $\OO_2^{\rm b}$, with $\xi_3\ne0$.
For an exchanged state $w\in\SSS_{0,\xi_3}$, define the BPZ pairing
\begin{equation}
    \tau_{12}(w;1,2)
    ={}_{\rm BPZ}\langle w\rvert
    \OO_1^{\rm b}(x_1,t_1)\OO_2^{\rm b}(x_2,t_2)\lvert0\rangle
    \, ,
    \qquad x_i=(u_i,z_i,\bz_i)\, .
    \label{eq:opeTauVacuum}
\end{equation}
We compute it first on the generating primary state $\lvert\OO_3(t)\rangle=\OO_3^{\rm b}(0,t)\lvert0\rangle$ of \eqref{eq:generatingstate}, and then on its descendants.
The constant $B_0$ retains the BPZ normalization in \eqref{eq:bpz}.
To compute the OPE about $x_2$, we use the translated frame $x_1=(u,z,\bz)=x_{12}$, $x_2=0$.

Apply the BPZ inversion \eqref{eq:opeBPZinversion} to the third insertion of $G_3$, transporting its residue prescription.
The outgoing primary coefficient at the origin of the inverted chart is
\begin{align}
    \left.(\mathcal{I}_{\xi_3}^{(3)}G_3)\right|_{x_3=0}
    &=C_3e^{\mathcal{E}_{12|3}}\delta_{\mathbb{C}}(\mathcal{C}_{12|3})
    \, ,
    \label{eq:opePrimaryCurrent}\\
    \mathcal{E}_{12|3}(t_3)
    &=\frac{u}{z}\left(\frac{\xi_3}{t_3}-\xi_1t_1-\xi_2t_2\right),\notag\\
    \mathcal{C}_{12|3}(t_3)
    &=z\left(\frac{\xi_1}{t_1}+\frac{\xi_2}{t_2}-\xi_3t_3\right)
    -\bz\left(\xi_1t_1+\xi_2t_2-\frac{\xi_3}{t_3}\right)
    \, .
    \label{eq:opePrimarySupport}
\end{align}
Here $z\ne0$ and $C_3=C_3(\xi_1,\xi_2,\xi_3)$ is the coefficient in \eqref{eq:threepointsolution}.
The inversion prefactor $1/(z_3\bz_3)$ cancels the holomorphic rescaling of the delta function, and the exponent becomes regular at the origin of the inverted chart on its transformed support.

Choose a simple nonzero root $t_3=t_3^\star$ of $\mathcal{C}_{12|3}=0$, or equivalently
\begin{equation}
    \xi_3z(t_3^\star)^2
    -\left[z\left(\frac{\xi_1}{t_1}+\frac{\xi_2}{t_2}\right)
    -\bz(\xi_1t_1+\xi_2t_2)\right]t_3^\star-\xi_3\bz=0
    \, .
    \label{eq:opePrimaryRoot}
\end{equation}
Continue this root holomorphically in the external data, avoiding zero and root collisions.
All residues below use its transported small positively oriented contour, enclosing neither $0$ nor the other root.

The formal delta of Appendix~\ref{app:deltaformal} collects the primary generating state, so the holomorphic pairing gives
\begin{align}
    \tau_{12}(\lvert\OO_3(t)\rangle;1,2)
    &=\int\delta^\times(t/t_3)
    \left.(\mathcal{I}_{\xi_3}^{(3)}G_3)\right|_{x_3=0}
    \frac{\dd t_3}{t_3}\notag\\
    &=\operatorname{Res}_{t_3=t_3^\star}
    \delta^\times(t/t_3)
    \frac{C_3e^{\mathcal{E}_{12|3}(t_3)}}{\mathcal{C}_{12|3}(t_3)}
    \frac{\dd t_3}{t_3}
    \label{eq:opeTauInvertedPairing}\\
    &=-\frac{C_3e^{\mathcal{E}_{12|3}(t_3^\star)}}
    {\xi_3(zt_3^\star+\bz/t_3^\star)}\,
    \delta^\times(t/t_3^\star)
    \, .
    \label{eq:opePrimaryTau}
\end{align}
Here $t$ labels the generating state, while $t_3$ is the residue variable of the inverted third insertion.
Each residue is understood coefficientwise in $t$.
The last equality uses $t_3\partial_{t_3}\mathcal{C}_{12|3}|_{t_3^\star} =-\xi_3(zt_3^\star+\bz/t_3^\star)$.
Only the third variable is paired.
Multiplying the result by $\dd t_1/t_1\wedge\dd t_2/t_2$ gives the remaining two-form.
Thus $t_1,t_2$ stay external; the two-cycle $\gamma$ in Appendix~\ref{app:deltaleray} is needed only for a further integration over them.

Contour deformations preserving the selected root and winding number leave this pairing unchanged.
Choosing the other root generally changes both the formal delta and its prefactor; the Ward equations do not select a branch.
A fixed linear combination of root contours gives the same combination of primary and descendant pairings.
We retain the chosen simple-root branch throughout.

To obtain the descendant pairings, the BPZ relation \eqref{eq:opeBPZdual} and $X\lvert0\rangle=0$ imply
\begin{equation}
    \begin{aligned}
    \tau_{12}(X^\dagger w;1,2)
    &={}_{\rm BPZ}\langle w\rvert
    [X,\OO_1^{\rm b}(x_1,t_1)\OO_2^{\rm b}(x_2,t_2)]\lvert0\rangle\\
    &=\bigl(\D_X^{{\rm b},(1)}+\D_X^{{\rm b},(2)}\bigr)
    \tau_{12}(w;1,2)
    \, .
    \end{aligned}
    \label{eq:opeInvertedWard}
\end{equation}
This identity holds before restriction to the translated frame.
For $X=L_+,P_{++}$, all vector-field and multiplication coefficients of $\D_X^{{\rm b},(2)}$ vanish at $x_2=0$ by \eqref{eq:locallowering}, so the restriction commutes with successive summed actions.
Applying \eqref{eq:opeInvertedWard} coefficientwise to the generating state gives, for $0\le b\le\ell$,
\begin{equation}
    \tau_{12}\bigl(L_-^bP_{--}^{\ell-b}\lvert\OO_3(t)\rangle;1,2\bigr)
    =\bigl(\D_{L_+}^{{\rm b},(1)}\bigr)^b
    \bigl(\D_{P_{++}}^{{\rm b},(1)}\bigr)^{\ell-b}
    \tau_{12}(\lvert\OO_3(t)\rangle;1,2)
    \, .
    \label{eq:opeTauDescendants}
\end{equation}
The order follows from $(L_-^bP_{--}^{\ell-b})^\dagger=P_{++}^{\ell-b}L_+^b$ and the reversed differential action of generator products; the two lowering operators commute.
The third null-state equation makes $\tau_{12}$ vanish on the null submodule, so these pairings are defined on the quotient.

Evaluate \eqref{eq:opeTauDescendants} by moving the external derivatives inside \eqref{eq:opeTauInvertedPairing}, with $t$ and $t_3$ fixed and the contour transported.
The formal delta is unaffected.
Define
\begin{equation*}
    \mathcal{H}_{12|3}(t_3)
    =\frac{\xi_1z}{t_1}-\xi_2\bz t_2+\frac{\xi_3\bz}{t_3}
    \, .
\end{equation*}
The explicit generators give, for integers $b,q\ge0$,
\begin{align}
    \D_{P_{++}}^{{\rm b},(1)}
    \left(\frac{z^b\mathcal{H}_{12|3}^{q}e^{\mathcal{E}_{12|3}}}
    {\mathcal{C}_{12|3}^{b+1}}\right)
    &=\frac{z^b\mathcal{H}_{12|3}^{q+1}e^{\mathcal{E}_{12|3}}}
    {\mathcal{C}_{12|3}^{b+1}},\notag\\
    \D_{L_+}^{{\rm b},(1)}
    \left(\frac{z^b\mathcal{H}_{12|3}^{q}e^{\mathcal{E}_{12|3}}}
    {\mathcal{C}_{12|3}^{b+1}}\right)
    &=(b+1)\frac{z^{b+1}\mathcal{H}_{12|3}^{q+1}e^{\mathcal{E}_{12|3}}}
    {\mathcal{C}_{12|3}^{b+2}}
    \, .
    \label{eq:opeWardDensityActions}
\end{align}
Repeatedly applying these identities to the integrand in \eqref{eq:opeTauInvertedPairing}, first $\ell-b$ times with $\D_{P_{++}}^{{\rm b},(1)}$ and then $b$ times with $\D_{L_+}^{{\rm b},(1)}$, gives the generating descendant pairing
\begin{equation}
    \tau_{12}\bigl(L_-^bP_{--}^{\ell-b}\lvert\OO_3(t)\rangle;1,2\bigr)
    =\operatorname{Res}_{t_3=t_3^\star}
    \delta^\times(t/t_3)T_{\ell,b}(t_3)\frac{\dd t_3}{t_3}
    \, ,
    \label{eq:opeGeneratingDescendants}
\end{equation}
where the explicit meromorphic factor is
\begin{equation}
    T_{\ell,b}(t_3)
    =C_3b!z^b\frac{\mathcal{H}_{12|3}(t_3)^\ell
    e^{\mathcal{E}_{12|3}(t_3)}}{\mathcal{C}_{12|3}(t_3)^{b+1}}
    \, .
    \label{eq:opeWardDensityClosed}
\end{equation}
This supplies all descendant pairings while retaining the auxiliary variable.
We take their component moments only when solving the OPE coefficient equations below.

\subsection{Explicit OPE coefficients}
\label{app:opedescendants}

To solve for the OPE coefficients, return to the quotient basis $v_{\ell r,a}=[L_-^aP_{--}^{\ell-a}e_{r-a}]$ of Section~\ref{subsec:radical}, at $J=0$ and $\xi=\xi_3$.
Here $\ell$ is the level and $r$ the eigenvalue of $L_0-\bL_0$.
The state--operator correspondence sends $v_{\ell r,a}$ to $\partial_z^a\partial_u^{\ell-a}\OO_{3,r-a}^{\rm b}(0)$.
We define the OPE coefficients by
\begin{equation}
    \left.\OO_1^{\rm b}(x_1,t_1)\OO_2^{\rm b}(x_2,t_2)
    \right|_{\SSS_{0,\xi_3}}
    \sim\sum_{\substack{\ell\ge0,\ r\in\mathbb{Z}\\0\le a\le\ell}}
    c_{\ell r,a}(x_{12};t_1,t_2)\,
    \partial_{z_2}^{a}\partial_{u_2}^{\ell-a}
    \OO^{\rm b}_{3,r-a}(x_2)
    \, .
    \label{eq:opeLocalChannel}
\end{equation}
The restriction denotes the contribution of $\SSS_{0,\xi_3}$; the formal expansion is tested by BPZ pairings with algebraic states.
No independent $\partial_{\bz}$ descendants occur, since the quotient relation \eqref{eq:reduction} eliminates them.
Below we suppress the arguments $(x_{12};t_1,t_2)$ of the coefficients.

Comparing the coefficients of $t^{r-b}$ in \eqref{eq:opeGeneratingDescendants} gives the pairings needed for this basis:
\begin{equation}
    T_{\ell r,b}\coloneqq\tau_{12}(v_{\ell r,b};1,2)
    =\operatorname{Res}_{t_3=t_3^\star}
    t_3^{b-r}T_{\ell,b}(t_3)\frac{\dd t_3}{t_3}
    \, ,
    \qquad 0\le b\le\ell\, .
    \label{eq:opeDescendantRecursion}
\end{equation}
Thus the component data follow directly from the generating result \eqref{eq:opeWardDensityClosed}.

The Gram matrix of this quotient basis is
\begin{equation}
    (\mathsf{B}_{\ell,r})_{ab}=\hat{B}(v_{\ell r,a},v_{\ell r,b})
    \, ,
    \qquad 0\le a,b\le\ell\, .
    \label{eq:opeQuotientGram}
\end{equation}
It is nondegenerate for $B_0\ne0$ and $\xi_3\ne0$ by \eqref{eq:quotientdet}.
Its entries are
\begin{equation}
    (\mathsf{B}_{\ell,r})_{ab}=
    \begin{cases}
        0,&a+b<\ell
        \, ,\\[3pt]
        B_0a!b!\,\xi_3^{\,2\ell-a-b}\displaystyle\binom{r}{a+b-\ell},
        &a+b\ge\ell
        \, .
    \end{cases}
    \label{eq:opeGramExplicit}
\end{equation}
We use $\binom{r}{k}=r(r-1)\cdots(r-k+1)/k!$ for $k\ge1$ and $\binom{r}{0}=1$, including negative integer $r$.
To obtain \eqref{eq:opeGramExplicit}, move the second set of descendants through the BPZ pairing.
The commutators with $L_-$ and $P_{--}$ and the primary action \eqref{eq:seed} give, for $d=a+b-\ell\ge0$,
\begin{align}
    P_{++}^{\ell-b}v_{\ell r,a}
    &=\frac{a!}{d!}\xi_3^{\ell-b}[L_-^dP_{--}^{b-d}e_{r-d}],\notag\\
    L_+^b[L_-^dP_{--}^{b-d}e_{r-d}]
    &=b!\xi_3^{b-d}(r-d+1)_d[e_{r-b}]
    \, .
    \label{eq:opeGramReduction}
\end{align}
The primary pairing supplies $B_0$, and $(r-d+1)_d/d!=\binom{r}{d}$, where $(x)_d$ is the rising factorial with $(x)_0=1$.
For $a+b<\ell$, the first lowering action vanishes.

Pairing \eqref{eq:opeLocalChannel} with the BPZ dual of $v_{\ell r,b}$ gives
\begin{equation}
    \sum_{a=0}^{\ell}(\mathsf{B}_{\ell,r})_{ab}c_{\ell r,a}
    =T_{\ell r,b}
    \, ,\qquad b=0,\ldots,\ell\, .
    \label{eq:opeCoefficientSystem}
\end{equation}
Relabeling the row by $b=\ell-a$ and substituting the Gram entries reduces this system to
\begin{equation}
    \sum_{j=0}^{\ell-a}\binom{r}{j}\xi_3^{-j}(a+j)!
    c_{\ell r,a+j}
    =\frac{T_{\ell r,\ell-a}}{B_0(\ell-a)!\xi_3^\ell}
    \, .
    \label{eq:opeExplicitRecursion}
\end{equation}
The binomial identity $(1+s)^r(1+s)^{-r}=1$ implies $\sum_{j=0}^{N}\binom{r}{j}\binom{-r}{N-j}=\delta_{N0}$, which inverts this finite convolution:
\begin{equation}
    c_{\ell r,a}
    =\frac{1}{B_0a!\xi_3^\ell}
    \sum_{j=0}^{\ell-a}\frac{\binom{-r}{j}}{\xi_3^j}
    \frac{T_{\ell r,\ell-a-j}}{(\ell-a-j)!}
    \, ,\qquad 0\le a\le\ell\, .
    \label{eq:opeCoefficientClosed}
\end{equation}
Together with \eqref{eq:opeWardDensityClosed} and \eqref{eq:opeDescendantRecursion}, this determines each coefficient by a finite sum of explicit residues.
It holds for every integer $r$; only $B_0$, $\xi_3$ and factorials occur in denominators.

For the simultaneous collision $(u,z,\bz)=\varepsilon(\upsilon,1,\vartheta)$ with fixed direction and auxiliary variables, $t_3^\star$ is independent of $\varepsilon$ and the primary pairing scales as $\varepsilon^{-1}$.
The OPE \eqref{eq:opeLocalChannel} is a local formal expansion away from degenerate directions and root collisions.
The spectrum, $C_3$, $B_0$ and simple-root prescription remain input data; the nondegenerate Gram matrix fixes the coefficients for these data but selects neither a physical branch nor a global pairing.
No analytic convergence is assumed.

\bibliographystyle{JHEP}
\bibliography{references}

\providecommand{\href}[2]{#2}\begingroup\raggedright\begin{thebibliography}{10}

\bibitem{CheungDeLaFuenteSundrum2017}
C.~Cheung, A.~de~la Fuente and R.~Sundrum, \emph{{4D scattering amplitudes and asymptotic symmetries from 2D CFT}}, \href{https://doi.org/10.1007/JHEP01(2017)112}{\emph{JHEP} {\bfseries 01} (2017) 112} [\href{https://arxiv.org/abs/1609.00732}{{\ttfamily 1609.00732}}].

\bibitem{PasterskiShaoStrominger2017}
S.~Pasterski, S.-H.~Shao and A.~Strominger, \emph{{Flat Space Amplitudes and Conformal Symmetry of the Celestial Sphere}}, \href{https://doi.org/10.1103/PhysRevD.96.065026}{\emph{Phys. Rev. D} {\bfseries 96} (2017) 065026} [\href{https://arxiv.org/abs/1701.00049}{{\ttfamily 1701.00049}}].

\bibitem{PasterskiShao2017}
S.~Pasterski and S.-H.~Shao, \emph{{Conformal basis for flat space amplitudes}}, \href{https://doi.org/10.1103/PhysRevD.96.065022}{\emph{Phys. Rev. D} {\bfseries 96} (2017) 065022} [\href{https://arxiv.org/abs/1705.01027}{{\ttfamily 1705.01027}}].

\bibitem{Strominger2018}
A.~Strominger, \emph{{Lectures on the Infrared Structure of Gravity and Gauge Theory}}, Princeton University Press (2018), [\href{https://arxiv.org/abs/1703.05448}{{\ttfamily 1703.05448}}].

\bibitem{Raclariu2021}
A.-M.~Raclariu, \emph{{Lectures on Celestial Holography}},  \href{https://arxiv.org/abs/2107.02075}{{\ttfamily 2107.02075}}.

\bibitem{Pasterski2021}
S.~Pasterski, \emph{{Lectures on celestial amplitudes}}, \href{https://doi.org/10.1140/epjc/s10052-021-09846-7}{\emph{Eur. Phys. J. C} {\bfseries 81} (2021) 1062} [\href{https://arxiv.org/abs/2108.04801}{{\ttfamily 2108.04801}}].

\bibitem{PasterskiPateRaclariu2021}
S.~Pasterski, M.~Pate and A.-M.~Raclariu, \emph{{Celestial Holography}},  in \emph{{Snowmass 2021}}, 2021 [\href{https://arxiv.org/abs/2111.11392}{{\ttfamily 2111.11392}}].

\bibitem{McLoughlinPuhmRaclariu2022}
T.~McLoughlin, A.~Puhm and A.-M.~Raclariu, \emph{{The SAGEX review on scattering amplitudes chapter 11: soft theorems and celestial amplitudes}}, \href{https://doi.org/10.1088/1751-8121/ac9a40}{\emph{J. Phys. A} {\bfseries 55} (2022) 443012} [\href{https://arxiv.org/abs/2203.13022}{{\ttfamily 2203.13022}}].

\bibitem{LawZlotnikov2020}
Y.T.A.~Law and M.~Zlotnikov, \emph{{Massive Spinning Bosons on the Celestial Sphere}}, \href{https://doi.org/10.1007/JHEP06(2020)079}{\emph{JHEP} {\bfseries 06} (2020) 079} [\href{https://arxiv.org/abs/2004.04309}{{\ttfamily 2004.04309}}].

\bibitem{IacobacciMuck2020}
L.~Iacobacci and W.~M{\"u}ck, \emph{{Conformal Primary Basis for Dirac Spinors}}, \href{https://doi.org/10.1103/PhysRevD.102.106025}{\emph{Phys. Rev. D} {\bfseries 102} (2020) 106025} [\href{https://arxiv.org/abs/2009.02938}{{\ttfamily 2009.02938}}].

\bibitem{Narayanan2020}
S.A.~Narayanan, \emph{{Massive Celestial Fermions}}, \href{https://doi.org/10.1007/JHEP12(2020)074}{\emph{JHEP} {\bfseries 12} (2020) 074} [\href{https://arxiv.org/abs/2009.03883}{{\ttfamily 2009.03883}}].

\bibitem{ChangMa2023}
C.-M.~Chang and W.-J.~Ma, \emph{{Missing corner in the sky: massless three-point celestial amplitudes}}, \href{https://doi.org/10.1007/JHEP04(2023)051}{\emph{JHEP} {\bfseries 04} (2023) 051} [\href{https://arxiv.org/abs/2212.07025}{{\ttfamily 2212.07025}}].

\bibitem{LiuMa2025}
R.~Liu and W.-J.~Ma, \emph{{Regular celestial amplitudes}},  \href{https://arxiv.org/abs/2512.05882}{{\ttfamily 2512.05882}}.

\bibitem{Donnay2022}
L.~Donnay, A.~Fiorucci, Y.~Herfray and R.~Ruzziconi, \emph{{Carrollian Perspective on Celestial Holography}}, \href{https://doi.org/10.1103/PhysRevLett.129.071602}{\emph{Phys. Rev. Lett.} {\bfseries 129} (2022) 071602} [\href{https://arxiv.org/abs/2202.04702}{{\ttfamily 2202.04702}}].

\bibitem{Bagchi2022}
A.~Bagchi, S.~Banerjee, R.~Basu and S.~Dutta, \emph{{Scattering Amplitudes: Celestial and Carrollian}}, \href{https://doi.org/10.1103/PhysRevLett.128.241601}{\emph{Phys. Rev. Lett.} {\bfseries 128} (2022) 241601} [\href{https://arxiv.org/abs/2202.08438}{{\ttfamily 2202.08438}}].

\bibitem{Nguyen2025}
K.~Nguyen, \emph{{Lectures on Carrollian Holography}},  \href{https://arxiv.org/abs/2511.10162}{{\ttfamily 2511.10162}}.

\bibitem{Banerjee2019}
S.~Banerjee, \emph{{Null Infinity and Unitary Representation of The Poincare Group}}, \href{https://doi.org/10.1007/JHEP01(2019)205}{\emph{JHEP} {\bfseries 01} (2019) 205} [\href{https://arxiv.org/abs/1801.10171}{{\ttfamily 1801.10171}}].

\bibitem{NguyenWest2023}
K.~Nguyen and P.~West, \emph{{Carrollian Conformal Fields and Flat Holography}}, \href{https://doi.org/10.3390/universe9090385}{\emph{Universe} {\bfseries 9} (2023) 385} [\href{https://arxiv.org/abs/2305.02884}{{\ttfamily 2305.02884}}].

\bibitem{Nguyen2024}
K.~Nguyen, \emph{{Carrollian conformal correlators and massless scattering amplitudes}}, \href{https://doi.org/10.1007/JHEP01(2024)076}{\emph{JHEP} {\bfseries 01} (2024) 076} [\href{https://arxiv.org/abs/2311.09869}{{\ttfamily 2311.09869}}].

\bibitem{ChenHu2024}
B.~Chen and Z.~Hu, \emph{{Bulk reconstruction in flat holography}}, \href{https://doi.org/10.1007/JHEP03(2024)064}{\emph{JHEP} {\bfseries 03} (2024) 064} [\href{https://arxiv.org/abs/2312.13574}{{\ttfamily 2312.13574}}].

\bibitem{ZhengDescendants2026}
Y.-f.~Zheng, \emph{{Missing Descendants in the Carrollian Conformal Family}},  \href{https://arxiv.org/abs/2607.28400}{{\ttfamily 2607.28400}}.

\bibitem{Zheng2026}
Y.-f.~Zheng, \emph{{Carrollian Dictionary for Massive Particles at Null Infinity}},  \href{https://arxiv.org/abs/2608.11945}{{\ttfamily 2608.11945}}.

\bibitem{Have2024}
E.~Have, K.~Nguyen, S.~Prohazka and J.~Salzer, \emph{{Massive carrollian fields at timelike infinity}}, \href{https://doi.org/10.1007/JHEP07(2024)054}{\emph{JHEP} {\bfseries 07} (2024) 054} [\href{https://arxiv.org/abs/2402.05190}{{\ttfamily 2402.05190}}].

\bibitem{Melton2026}
W.~Melton, K.~Michaelsen and R.~Ruzziconi, \emph{{Observing Massive Scattering from Null Infinity}},  \href{https://arxiv.org/abs/2606.27421}{{\ttfamily 2606.27421}}.

\bibitem{Liu2026}
R.~Liu, W.-J.~Ma, H.~Zheng and Y.-f.~Zheng, \emph{{Carrollian holography with agentic AI: Real mass is imaginary}},  \href{https://arxiv.org/abs/2606.05401}{{\ttfamily 2606.05401}}.

\bibitem{Atanasov2021}
A.~Atanasov, W.~Melton, A.-M.~Raclariu and A.~Strominger, \emph{{Conformal block expansion in celestial CFT}}, \href{https://doi.org/10.1103/PhysRevD.104.126033}{\emph{Phys. Rev. D} {\bfseries 104} (2021) 126033} [\href{https://arxiv.org/abs/2104.13432}{{\ttfamily 2104.13432}}].

\bibitem{ChenHuMao2026}
B.~Chen, Z.~Hu and X.-C.~Mao, \emph{{QFT in Klein space}}, \href{https://doi.org/10.1103/y1sk-kh72}{\emph{Phys. Rev. D} {\bfseries 113} (2026) 085005} [\href{https://arxiv.org/abs/2505.16436}{{\ttfamily 2505.16436}}].

\end{thebibliography}\endgroup
\end{document}